\documentclass[a4paper,fleqn,usenatbib]{mnras}
\usepackage{newtxtext,newtxmath}
\usepackage[T1]{fontenc}
\usepackage{ae,aecompl}

\usepackage{amsmath}	% Advanced maths commands
\usepackage{amssymb}	% Extra maths symbols
\usepackage{bm}
\usepackage{here}
\usepackage[pdftex]{graphicx}

\usepackage{color}

\title[The thermal-radiative wind in GX 13+1]
{The thermal-radiative wind in the neutron star low mass X-ray binary GX 13+1 
}

\author[Tomaru et al.]{
\thanks{E-mail: ryota.tomaru@ipmu.jp}
Ryota Tomaru,$^{1}$
Chris Done,$^{3,1}$
Ken Ohsuga,$^{4}$
Hirokazu Odaka$^{2,1}$ and
\newauthor
Tadayuki Takahashi$^{1,2}$ 
\\
$^{1}$Kavli Institute for the Physics and Mathematics of the Universe (WPI), The University of Tokyo, Kashiwa 277-8583, Japan  \\
$^{2}$Department of Physics, The University of Tokyo, 7-3-1 Hongo, Bunkyo, Tokyo 113-0033, Japan\\
$^{3}$ Centre for Extragalactic Astronomy, Department of Physics, Durham University, South Road, Durham, DH1 4ED, UK\\
$^{4}$Center for Computational Sciences, University of Tsukuba, 1-1-1- Ten-nodai, Tsukuba, Ibaraki, 305-8577, Japan \\
}

\date{Accepted XXX. Received YYY; in original form ZZZ}

\pubyear{2019}

\begin{document}
\label{firstpage}
\pagerange{\pageref{firstpage}--\pageref{lastpage}}
\maketitle

\begin{abstract}

We fit the observed high ionisation X-ray absorption lines in the neutron star binary GX13+1 with a full simulation of a thermal-radiative wind.
This uses a radiation hydrodynamic code coupled to  Monte Carlo radiation transfer to compute the observed line profiles from Hydrogen and Helium-like iron and Nickel,
including all strong K$\alpha$ and K$\beta$ transitions.
The wind is very strong as this object has a very large disc and is very luminous. The absorption lines from Fe K$\alpha$ are strongly saturated as the ion columns are large, 
so the line equivalent widths (EWs) depend sensitively on the velocity structure. 
We additionally simulate the lines including isotropic turbulence at the level of the azimuthal and radial velocities. 
We fit these models to the Fe {\sc xxv} and {\sc xxvi} absorption lines seen in the highest resolution Chandra third order HETGS data.
These data already rule out the addition of turbulence at the level of the radial velocity of $\sim 500$ km/s.
The velocity structure predicted by the thermal-radiative wind alone is a fairly good match to the observed profile, with an upper limit to additional turbulence at the level of the azimuthal velocity of $\sim 100$ km/s.
This gives stringent constraints on any remaining contribution from magnetic acceleration.
\end{abstract}

\section{introduction}

Accretion disc winds are clearly present in the black hole and neutron star binaries, with blueshifted, highly ionised (predominantly Fe {\sc xxv} and {\sc xxvi} K$\alpha$) absorption lines detected in multiple systems (e.g. \citealt{Ponti2012, DiazTrigo2016}).
These give us a small scale version of accretion disc winds in supermassive black holes which drive AGN feedback (e.g. \citealt{Nardini2015}). 
The important questions in both systems are how the winds are launched and accelerated - whether the forces are magnetic or thermal/radiative.
Either one of these has important consequences.
Magnetic fields link to the jet, and also to the mechanism for accretion itself as they transport angular momentum outwards, allowing the material to accrete.
Alternatively, thermal-radiative winds can eject more mass from the outer disc than accretes through the inner disc, controlling the amount of material available to power the system. 

The galactic binaries are more easily studied than the AGN as they are typically much brighter, with enough signal-to-noise to give well exposed data from the 
Chandra high energy transmission grating spectrometer (HETGS), the 
instrument with the best current energy resolution around the iron lines. 
They also form a fairly homogeneous population, so the properties of the wind in different sources show how the wind depends on geometrical factors such as disc size , inclination, and physical properties such as X-ray luminosity and spectral shape.
These studies clearly show that wind absorption lines are seen in a very specific subset of the disc accreting binary systems.
They are all high inclination systems, indicating that wind geometry is equatorial \citep{Ponti2012}.
More surprisingly, they are seen only when these systems have strong inner disc emission (soft states in black hole binaries, banana branches in neutron stars),
and disappear when the same source makes a spectral transition to a harder state (termed low/hard states in black holes, or island branches in neutron stars)  \citep{Ponti2012, Ponti2014}.
This behaviour of the absorption features is opposite to that of the compact radio jet, which is seen in the hard state and disappears in the soft state. 
%The disappearance of the absorption features as the source changes from soft to hard states is opposite to the behaviour of the compact radio jet which is seen in the hard states and disappears in the soft.
This anti-correlation of jet and wind led to speculation that these were the same magnetically powered outflow,
changing its geometry at the spectral state transition \citep{Neilsen2009, Miller2012}.

%However, the observed wind outflow velocities are rather slow, a few hundred to a few thousand km/s.
Magnetic driving is almost certainly important in jet acceleration \citep{Lynden-Bell1996,Blandford1977}, but the mechanism(s) required for winds 
are not so clear.
Unlike relativistic jets, the observed wind outflow velocities are rather slow, a few hundred to a few thousand km/s.
This velocity indicates they are launched from the outer disc since winds have terminal velocities which are similar to the escape velocity from their launch point,
whereas the jet surely is launched from the very inner regions.
Another feature which points to an outer disc origin for winds is that they are preferentially seen in long orbital period systems,
i.e. where the disc is large \citep{DiazTrigo2016}. 
Those features are consistent with thermal winds, which is driven by the radiative heating of illumination from the central source.
The central source illuminates the disc surface, heating it to the radiation (Compton) temperature, $T_\text{IC}$.
This temperature is constant at all radii, but gravity decreases outwards, so this gas temperature gives particle velocities which are higher than an escape for $kT_\text{IC}=GMm_p/R_\text{IC}$, defining the Compton radius $R_\text{IC}\sim 6.4\times 10^4/T_\text{IC,8} R_g$, where $R_g=GM/c^2$ and $T_\text{IC,8} = T_\text{IC}/10^8$ which is typically 0.1-0.2 in the soft states \citep{Begelman1983a, Done2018}.
These thermal winds are given an extra push by radiation pressure (thermal-radiative winds) when $L/L_\text{Edd}>0.2$, helping them escape. 

Thy dynamics of thermal-radiative winds can be calculated from other observational information such as the spectral shape, the luminosity,  and the disc size, %first principles, 
unlike the magnetic winds which depend on the unknown magnetic field configuration. 
Hence the thermal-radiative wind models are predictive, 
whereas the current magnetic wind models \citep{Chakravorty2016, Fukumura2017}, which are driven by magnetic fields threading accretion discs as represented by \citet{Blandford1982}, are only postdictive, 
constraining the magnetic configuration from the observed wind features.
We have developed a state-of-the-art set of codes to predict the thermal-radiative wind and its absorption/emission features.
This uses a full radiation-hydrodynamic simulation to predict the wind density and velocity from a given spectrum, luminosity and disc size \citep[here after T19]{Tomaru2019},
then uses this as input into a Monte Carlo radiation transfer code to predict the detailed line emission/absorption profile \citep[hereafter T18 and T20]{Tomaru2018, Tomaru2020}. 
Our results so far for the black hole binary H1743-322 show that thermal-radiative winds can match the best current data from the Chandra HETGS in the soft state,
and explain the wind disappearance in the hard state of this same object (T19).
These simulations also demonstrate that radiation pressure is an important contributor to the velocity profile (T19), 
making them different to the only other modern hydrodynamic simulation of thermal winds (e.g. \citealt{Luketic2010, Higginbottom2015, Higginbottom2018}) which have not included radiation pressure.
The most recent simulation which includes radiation pressure also shows that effect of radiation force \citep{Higginbottom2020}.
These results shows radiation force is important to produce observed velocity even in X-ray binaries.

In this paper, we apply our full thermal-radiative model to the bright neutron star GX 13+1 which has the largest disc of any known galactic binary system in terms of gravitational radii, 
$R_\text{disc}\sim 5\times 10^6R_g$ (where $R_\text{g}=GM/c^2$) because of its long orbital period of 24 days and the relatively low mass of $1.4 M_\odot$. 
By comparison, the longest orbital period black hole, GRS 1915+105 (33.5 days), has a larger absolute disc size,
but its larger mass means that this corresponds to only $R_\text{disc}\sim  10^6~R_g$.
Large disc sizes are only possible with evolved donor stars, which have large mass transfer rates and hence produce highly luminous accretion flows.
The combination of large luminosity with a large disc predicts strong thermal (and thermal-radiative) winds (D18, T19a). 

We predict the detailed line profiles for GX13+1 from radiation transfer through a tailored radiation hydrodynamic simulation.
We fit these models to the Fe {\sc xxv} and {\sc xxvi} absorption lines seen in the highest resolution Chandra third order HETGS data (T18),
showing that these capture the main physical properties of the observed winds. The data strongly limit the amount of velocity shear and/or turbulence in the wind, which gives stringent constraints on a magnetic wind contribution.

\begin{table}
    \centering
    \caption{Detailed parameters for each line included in these {\sc monaco} simulations. 
    Note that we list only lines which have oscillator strength larger than $10^{-3}$.
    These are listed by increasing energy of the transition. 
    }
    \begin{tabular}{ccc}
    \hline
        Line ID &  Energy [keV] & Oscillator strength \\ 
        \hline
        Fe {\sc xxv} He$\alpha~y$    & 6.668 & $6.57 \times 10^{-2}$\\
        Fe {\sc xxv} He$\alpha~w$    & 6.700 & $7.26 \times 10^{-1}$\\
        Fe {\sc xxvi} Ly$\alpha_2$   & 6.952 & $1.36 \times 10^{-1}$\\
        Fe {\sc xxvi} Ly$\alpha_1$   & 6.973 & $2.73 \times 10^{-1}$\\
        Ni {\sc xxvii} He$\alpha~y$  & 7.765 & $8.50 \times 10^{-2}$\\
        Ni {\sc xxvii} He$\alpha~w$  & 7.805 & $7.06 \times 10^{-1}$\\
        Fe {\sc xxv} He$\beta~y$     & 7.872 & $1.37 \times 10^{-2}$\\
        Fe {\sc xxv} He$\beta~w$     & 7.881 & $1.39 \times 10^{-1}$\\
        Ni {\sc xxviii} Ly$\alpha_2$ & 8.073 & $1.36 \times 10^{-1}$\\
        Ni {\sc xxviii} Ly$\alpha_1$ & 8.102 & $2.72 \times 10^{-1}$\\
        Fe {\sc xxvi} Ly$\beta_2$   & 8.246 & $2.55 \times 10^{-2}$\\
        Fe {\sc xxvi} Ly$\beta_1$   & 8.253 & $5.23 \times 10^{-2}$\\
        Fe {\sc xxv} He$\gamma~y$   & 8.292 & $5.18 \times 10^{-3}$\\
        Fe {\sc xxv} He$\gamma~w$   & 8.295 & $5.10 \times 10^{-2}$\\
        \hline
    \end{tabular}
    \label{tab:line id}
\end{table}

\section{The NS LMXB GX 13+1: system parameters}
GX 13+1 is a bright, persistent NS LMXB in the Galactic bulge. 
This object orbits an evolved late-type K5 III star with $M_c = 5~M_\odot$ and its distance is estimated as $7\pm 1 \text{kpc}$ \citep{Bandyopadhyay1999}.
The infrared and X-ray light curves show that a binary orbital period is 24.5 days \citep{Corbet2010, Iaria2013}. 
There are X-ray dips \citep{D'Ai2014}, so this object has a moderately high inclination angle \citep[$60^\circ-80^\circ$, but more likely towards the lower end of this range as the dips are occasional events rather than seen regularly in each orbit]{DiazTrigo2012}.

The first moderate resolution spectra from {\it ASCA} \citet{Ueda2001} showed 
highly ionised iron K absorption lines, the first seen 
in a NS rather than BH binary system.
This absorption is most clearly seen in high resolution spectroscopy with {\it Chandra}/HETGS, where K$\alpha$ lines from H-like, Fe, Mn, Cr, Ca, Ar, S, Si, and Mg are detected as well as He-like Fe. The blueshift of these lines show that the material is outflowing with $v_\text{out}\sim 400~\text{km/s}$ \citep{Ueda2004,Madej2014,Allen2018}.

\begin{figure}
    \centering
    \includegraphics[width=0.9\hsize]{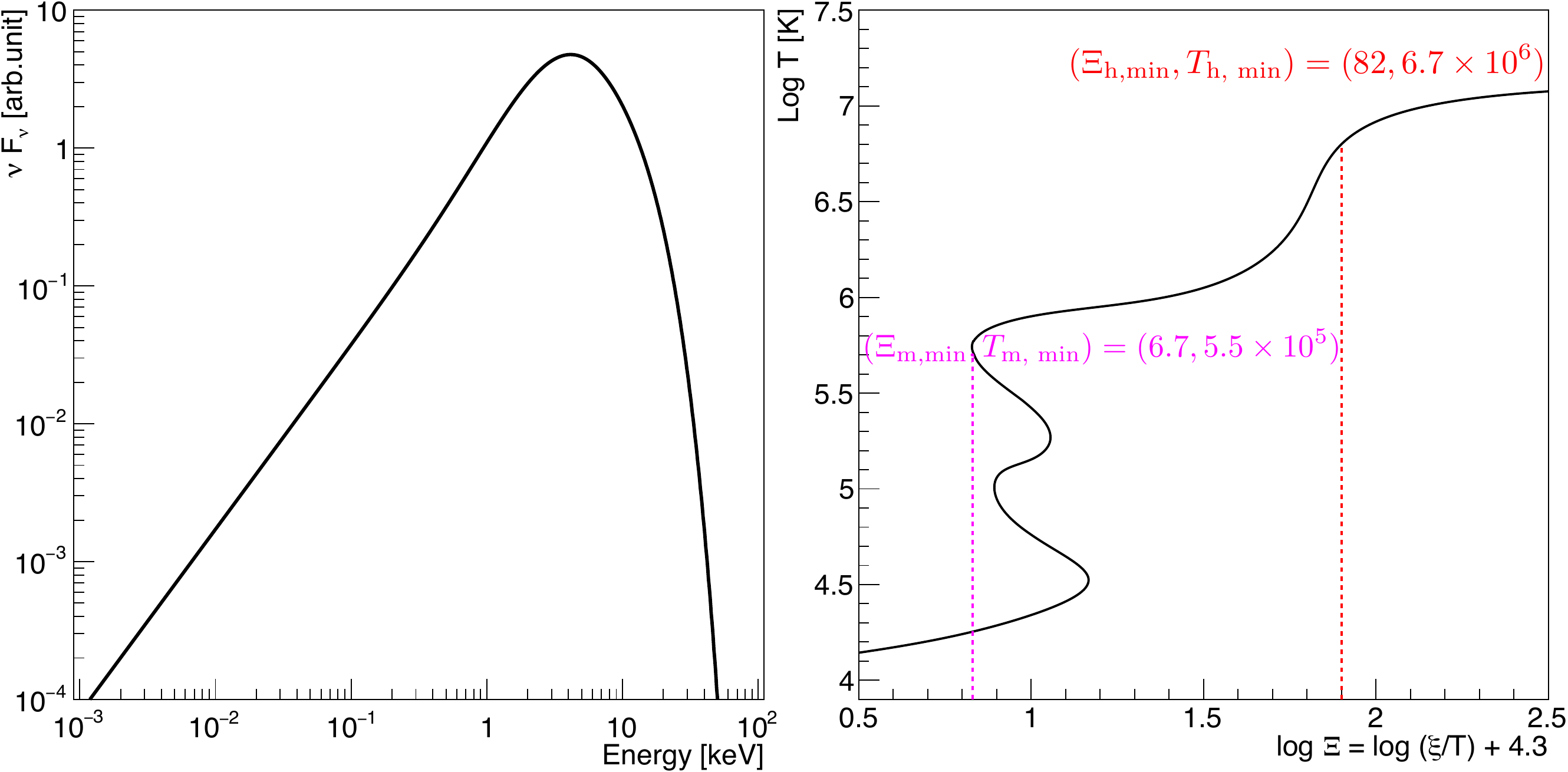}
    \caption{The SED of GX 13+1 determined from RXTE (left) and its thermal equilibrium curve (right). }
    \label{fig:GX13p1_SED}
\end{figure}

From the orbital period, we calculate the semi-major axis $a = 4.6\times 10^{12}~\mathrm{cm}$ and 
the Roche lobe radius $R_\text{RL} = 1.3 \times 10^{12}~\mathrm{cm}$, thus an approximate disc size is $R_\text{disc} = 1.0\times 10^{12}~\mathrm{cm}$  
($80\%$ of the Roche-lobe radius), which corresponds  to $5\times 10^{6}~R_g$.
The Compton radius with $T_\text{IC} = 1.3 \times 10^7~\text{K}$ is $R_\text{IC} = 5.0\times 10^{5}~R_g$, so the outer disc radius is $R_\text{disc} = 10~R_\text{IC}$.

\section{Observational data}

\begin{figure*}
    \centering
    \includegraphics[width=0.9\hsize]{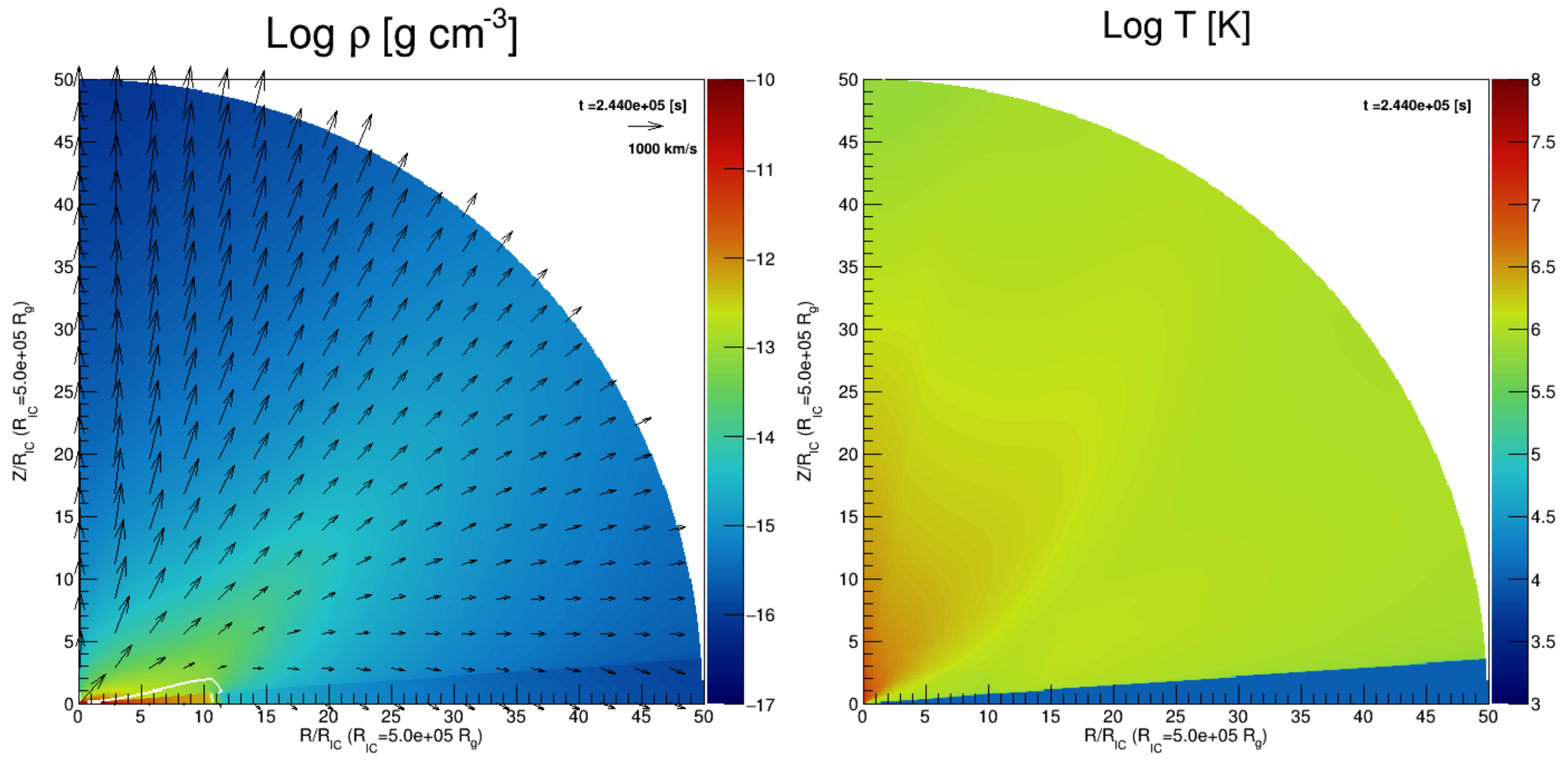}
    \caption{The radiation hydrodynamic simulation results with $L = 0.5 L_\text{Edd}$. The left panel shows the distribution of density, with velocity vectors while the right hand panel shows temperature.
    The white solid line shows Mach1 surface.
    }
    \label{fig:GX13p1_hydro}
\end{figure*}

\begin{table}
    \centering
    \caption{List of the {\it Chandra}/HETGS observations}
    \begin{tabular}{cccc}
    \hline 
    OBSID & MODE & Date & Exposure (ks) \\
    \hline 
    11815 &  TE & 24/07/2010 &28\\
    11816 &  TE & 30/07/2010 &28\\
    11814 &  TE & 01/08/2010 &28\\
    11817 &  TE & 03/08/2010 &28\\
    11818 &  CC & 05/08/2010 &24\\
    \hline
    \end{tabular}
    \label{tab:obs_list_GX13p1}
\end{table}

\if0
\begin{table}
    \centering
    \caption{Fits to the absorption lines of  OBSID: 11818 using {\sc kabs}.  
    Errors are the $90\% $ confidence level ($\Delta\chi^2=2.7$).}
    \begin{tabular}{cccc}
    \hline 
   ions  & $N_\mathrm{ion}~ [10^{18}~\mathrm{cm^{-2}}]$  & $kT~[\text{keV}]$ & $z\times 10^{-3}$ \\ 
   \hline
   %Fe \scriptsize{XXVI} & $3.6^{+46}_{-2.4}$ & $< 49$  & $ -0.91^{+0.37}_{-0.52}$\\
   Fe {\sc xxvi} & $3.6^{+46}_{-2.4}$ & $< 49$  & $ -0.9^{+0.4}_{-0.5}$\\
   Fe {\sc xxv} & $0.5^{+1.2}_{-0.2}$ & $7^{+14}_{-6}$  & $ -1.1^{+0.2}_{-0.6}$\\
   Ca {\sc XX} & $0.06^{+2.0}_{-0.02}$ & $3.8^{+11}_{-3.7}$  & $ -1.8^{+0.16}_{-0.7}$\\
   S {\sc xxvi} & $0.07 \pm 0.03 $ & $0.7^{+0.2}_{-0.4}$  & $ -1.27^{+0.06}_{-0.04}$\\
   %Si \scriptsize{XIV} & $0.052^{+0.040}_{-0.019}$ & $<7.5$  & $ -1.2^{+0.36}_{-0.11}$\\
   Si {\sc xiv} & $0.05 \pm 0.01 $ & $1.1^{+0.5}_{-0.3}$  & $ -1.2^{+0.07}_{-0.1}$\\
   Mg {\sc xii} & $0.014^{+0.8}_{-0.013}$ & $< 4.7$  & $ -1.0^{+0.14}_{-0.28}$\\
   \hline 
    \end{tabular}
    \label{tab:GX13p1_obs}
\end{table}
\fi
\begin{table}
    \centering
    \caption{Fits to the absorption lines of  OBSID: 11818 using {\sc kabs}.  
    Errors are the $90\% $ confidence level ($\Delta\chi^2=2.7$).}
    \scalebox{0.8}{
    \begin{tabular}{cccc}
    \hline 
   ions  & $N_\mathrm{ion}~ [10^{18}~\mathrm{cm^{-2}}]$  & $kT~[\text{keV}]~(\sqrt{2kT/m_\text{ion}}~ [\text{km/s}])$ & $z\times 10^{-3}$ \\ 
   \hline
   %Fe \scriptsize{XXVI} & $3.6^{+46}_{-2.4}$ & $< 49$  & $ -0.91^{+0.37}_{-0.52}$\\
   Fe {\sc xxvi} & $3.6^{+46}_{-2.4}$ & $< 49~ (<410)$ & $ -0.9^{+0.4}_{-0.5}$\\
   Fe {\sc xxv} & $0.5^{+1.2}_{-0.2}$ & $7^{+14}_{-6}~(160^{+160}_{-70})$  & $ -1.1^{+0.2}_{-0.6}$\\
   Ca {\sc XX} & $0.06^{+2.0}_{-0.02}$ & $3.8^{+11}_{-3.7}~(140^{+200}_{-70})$  & $ -1.8^{+0.16}_{-0.7}$\\
   S {\sc xxvi} & $0.07 \pm 0.03 $ & $0.7^{+0.2}_{-0.4}~(65^{+9}_{-18})$  & $ -1.27^{+0.06}_{-0.04}$\\
   %Si \scriptsize{XIV} & $0.052^{+0.040}_{-0.019}$ & $<7.5$  & $ -1.2^{+0.36}_{-0.11}$\\
   Si {\sc xiv} & $0.05 \pm 0.01 $ & $1.1^{+0.5}_{-0.3}~(87^{+20}_{-11})$  & $ -1.2^{+0.07}_{-0.1}$\\
   Mg {\sc xii} & $0.014^{+0.8}_{-0.013}$ & $< 4.7 (<190)$  & $ -1.0^{+0.14}_{-0.28}$\\
   \hline 
    \end{tabular}
    }
    \label{tab:GX13p1_obs}
\end{table}
There are 8 separate {\it Chandra}/HETGS observations, each showing highly ionised absorption lines \citep{Ueda2004,Allen2018}. 
We focus on the sequential observations in 2010 (Tab.\ref{tab:obs_list_GX13p1}). 
These all have similar continuum shape, and the EWs of the absorption lines are consistent within statistical errors for all these spectra \citep{Allen2018}. 
We use the first order spectra from the Continuous Clocking (CC) mode observation (ObsID 11818) to determine ion column densities from absorption lines, as this mode is less affected by pileup.
We extract the third order spectra from the remaining 4 observations and co-add them to increase the signal-to-noise (see T18).

The absorption line EW increases with increasing column density in the ion but saturates when the core of the line goes black. 
The line depth cannot increase below zero intensity, so 
increasing the ion absorption column beyond this point only makes the line slightly wider - its EW remains approximately constant.
This makes it difficult to unambiguously determine the ion column from the absorption line EW unless the line profile is fully resolved.

We use the {\sc kabs} absorption line model \cite{Ueda2004} to fit the data.
This model calculates the full Voigt profile rather than using a Gaussian function, giving an estimate for the ion column, which includes the effect of saturation.
This is why the uncertainties on the ion columns are large, even when the uncertainty of the line EW is small \citep{Ueda2004,Allen2018}.

We rewrite the {\sc kabs} model so that it runs faster, using the algorithm of \citet{Wells1999} to compute the Voigt function rather than using a numerical convolution. 
We also update the atomic database, to be consistent with that used in our radiation transfer code in Tab.\ref{tab:line id}.
We make this new version is publicly available as a local model for {\sc xspec}.

Tab.~\ref{tab:GX13p1_obs} shows the ion columns  ($N_\mathrm{ion}$) determined from the main absorption lines 
along with their Doppler width (expressed as ion temperature $kT$, assuming that all of the velocity widths are from this)
and outflow velocity (expressed as a blueshift $z = v/c$. Negative values mean blueshift). 
The ion temperature in the wind from our radiation hydrodynamic simulation is at most $T_\text{IC}\sim 1.4$~keV (see below) and  almost all observed temperature determined by absorption lines include that temperature obtained by our hydrodynamic simulation.
%All the rest have upper limits on the width which are much larger, where bulk motion rather than temperature dominates. 
We caution that these line widths can be affected by the distortion of the absorption line profile by a contribution from the same ion in emission, as the wind will have a P Cygni profile rather than just an absorption line.
Nonetheless, these give a baseline set of ion columns to compare with the results of the radiation hydrodynamic simulations. 

\section{Radiation hydrodynamic simulation}

\begin{figure}
    \centering
    \includegraphics[width=0.9\hsize]{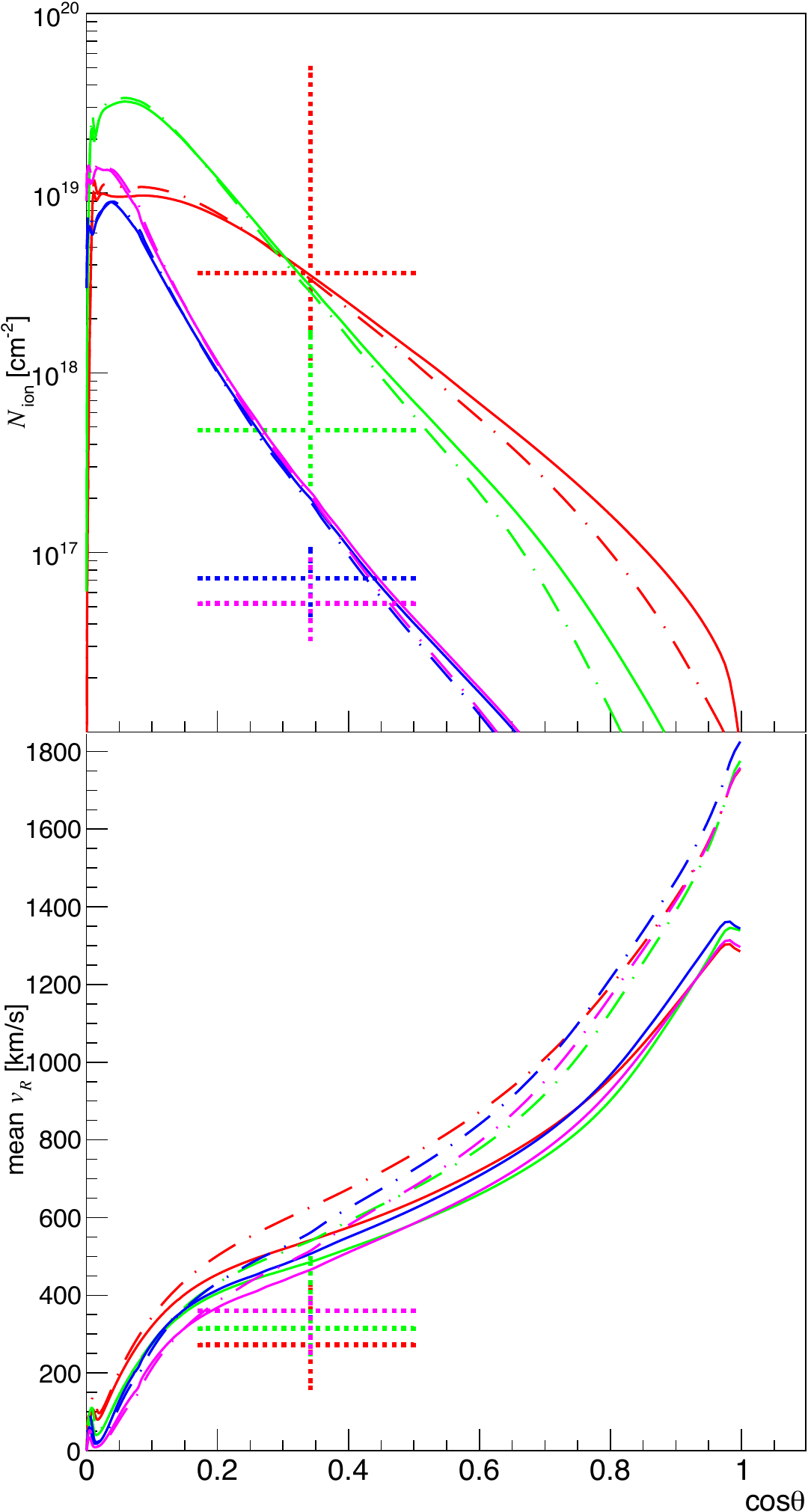}
    \caption{The angular dependence of ion columns (top) and ion column weighted mean velocities (bottom).
    The data points with errors are shown as dashed, to compare to the simulation results  with $L=0.5L_\text{Edd}$ (solid) and  $L=0.7L_\text{Edd}$ (dashed-dotted). 
    Each transition is shown in a different colour, Fe {\scriptsize XXVI} (red), Fe {\scriptsize XXV} (green), S {\scriptsize XVI} (blue), and Si {\scriptsize XIV} (magenta).
    }
    \label{fig:GX13p1_column}
\end{figure}

\begin{figure}
    \centering
    \includegraphics[width=0.9\hsize]{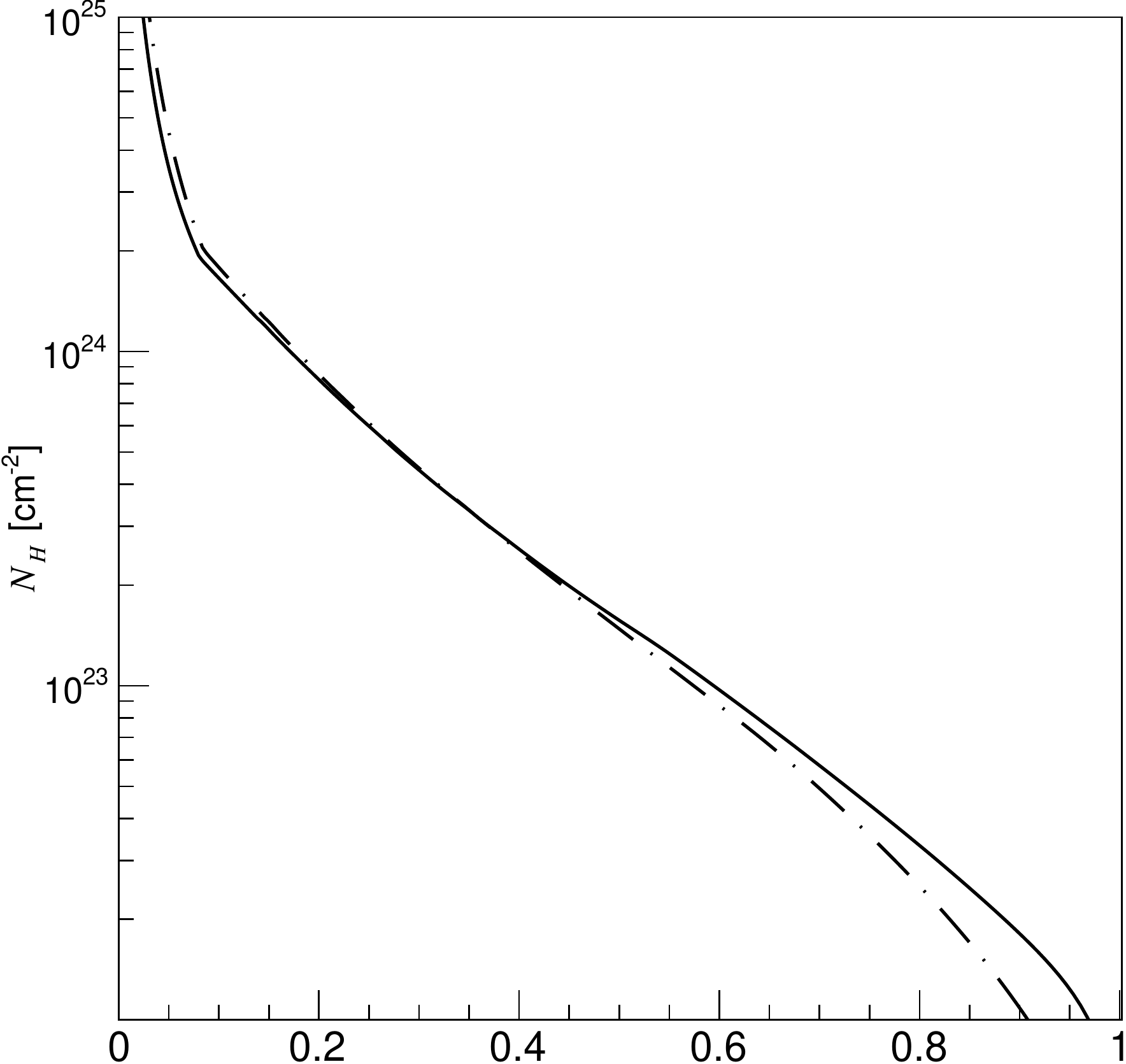}
     \caption{The angular dependence of total hydrogen column density in the radiation hydrodynamic simulations for  $(L/L_\text{Edd}=0.5$ (solid) and 
    $0.7$ (dash-dotted line).
    }
    \label{fig:NH}
\end{figure}

\subsection{Fiducial model: $L/L_\text{Edd}=0.5$}

Our radiation hydrodynamic code solves the same equations as in T19.
This code simulates the dynamics of winds by considering the effects of radiative heating, cooling and acceleration of material irradiated by the observed X-ray spectral energy distribution (SED) in a rotationally axisymmetric, 2 dimensional spherical polar coordinate system.
The heating and cooling include free-free bound-free, free-bound, bound-bound, and Compton processes, and the radiation force includes the bound-bound, bound-free and electron scattering.
At each time step, the net heating/cooling rates and radiation force is updated,
with values dependent on the computed ionisation parameter $\xi = L_x/(n_p R^2)$ (where $L_x$ is the luminosity corrected for attenuation from all sources of opacity and scattering along the line of sight)
and temperature of the gas. These rates are 
pre-calculated using {\sc cloudy} from the observed SED.

We extract the obseved SED from {\it RXTE}  data (OBSID: 95338-01-01-07) which is simultaneous with the CC observation of {\it Chandra}/HETGS (OBSID:11818).
The Compton temperature of this SED is $T_\text{IC} = 1.3\times 10^{7} \mathrm{K}$  and the corresponding Compton radius $R_\text{IC} = 5.0\times 10^5 R_g$, and its luminosity is $L\sim 0.5L_\text{Edd}$. 
We use these values for our fiducial simulation. 

While thermal winds are launched from the outer disc, 
T19 also highlight the importance of irradiation of the inner disc.
\cite{Begelman1983b} show that this results in a static X-ray heated atmosphere at small radii which quickly goes optically thick to electron scattering along the equatorial plane.
This atmosphere forms a shadow on the disc, shielding it from direct irradiation until the convex intrinsic shape of the disc lifts its surface above this shadow.
For the fiducial parameters, the radius at which the disc re-emerges from the inner region shadow is $R_\text{is}=0.2~R_\text{IC}$, where the optical depth of inner corona is unity.
Thus we choose a radial grid
for our radiation hydrodynamic simulation from $R_\text{in} = 0.05~R_\text{IC}$ and  $R_\text{out} =50~R_\text{IC}$ 
so as to resolve the entire region of the outer disc which is directly illuminated, and to capture the wind behaviour as it expands out beyond the outer edge of the disc at $10R_\text{IC}$. 

We calculate the thermal equilibrium curve from this observed SED
using {\sc cloudy} (right panel of Fig.\ref{fig:GX13p1_SED}).
This curve is very similar to that derived for the soft state SED in the black hole binary H1743-322 in that 
it has a complex shape rather than the simple S curve characteristic of harder spectra (see, e.g. Fig 9b in T19 and Fig 2b of \citep{Woods1996}). 
Similarly to H1743-322 (T19), we take the disc photosphere/wind boundary to be the minimum value of the pressure ionisation parameter ($\Xi = L/(4 \pi c R^2 nkT) = \xi/(4\pi c kT)$) on the middle branch,
$\Xi_\text{m,min}=6.7$, which has $T_\text{m, min} = 5.5 \times 10^5~\text{K}$. 
These values correspond a standard ionisation parameter of $\xi_\text{m,min} = 190$.
The material with this ionisation parameter already has fairly large opacity, so effectively shields material on the lower branches from irradiation.

We run the radiation hydrodynamic code of T19 using this SED.
We end the simulation after 10 sound crossing times of the disc radius ($10\times 10 R_\text{IC}/c_\text{IC} = 2.4 \times 10^5$ s, $c_\text{IC} = \sqrt{kT_\text{IC}/(0.61 m_p)}$) to ensure that this has converged,
and show an overview of the wind properties in Tab. ~\ref{tab:summary_GX13p1}.

The efficiency of the wind, $\eta=\dot{M}_\text{w}/\dot{M}_\text{a}\sim 8$.
This is about 4 times larger than in the analytic estimates of \citealt{Done2018} (cyan line in their 
Fig. 3 has the same disc size and Compton temperature as here and tends to $\eta\to 2$ at high luminosities).
This is due to their assumption that the base of the wind/photosphere of the disc was at $\Xi_\text{c,max}=40$.
The wind mass loss rate is set by the density at the base of the wind, which is inversely proportional to the pressure ionisation parameter at this point.
Hence using $\Xi_\text{m,min}=6.7$ rather than $\Xi_\text{c,max}=40$ increases the mass loss rate by this factor.
Thus the analytic estimates are a relatively good description of the simulation results when the correct heating/cooling is included.

\if0
\begin{figure*}
    \centering
    \includegraphics[width=0.45\hsize]{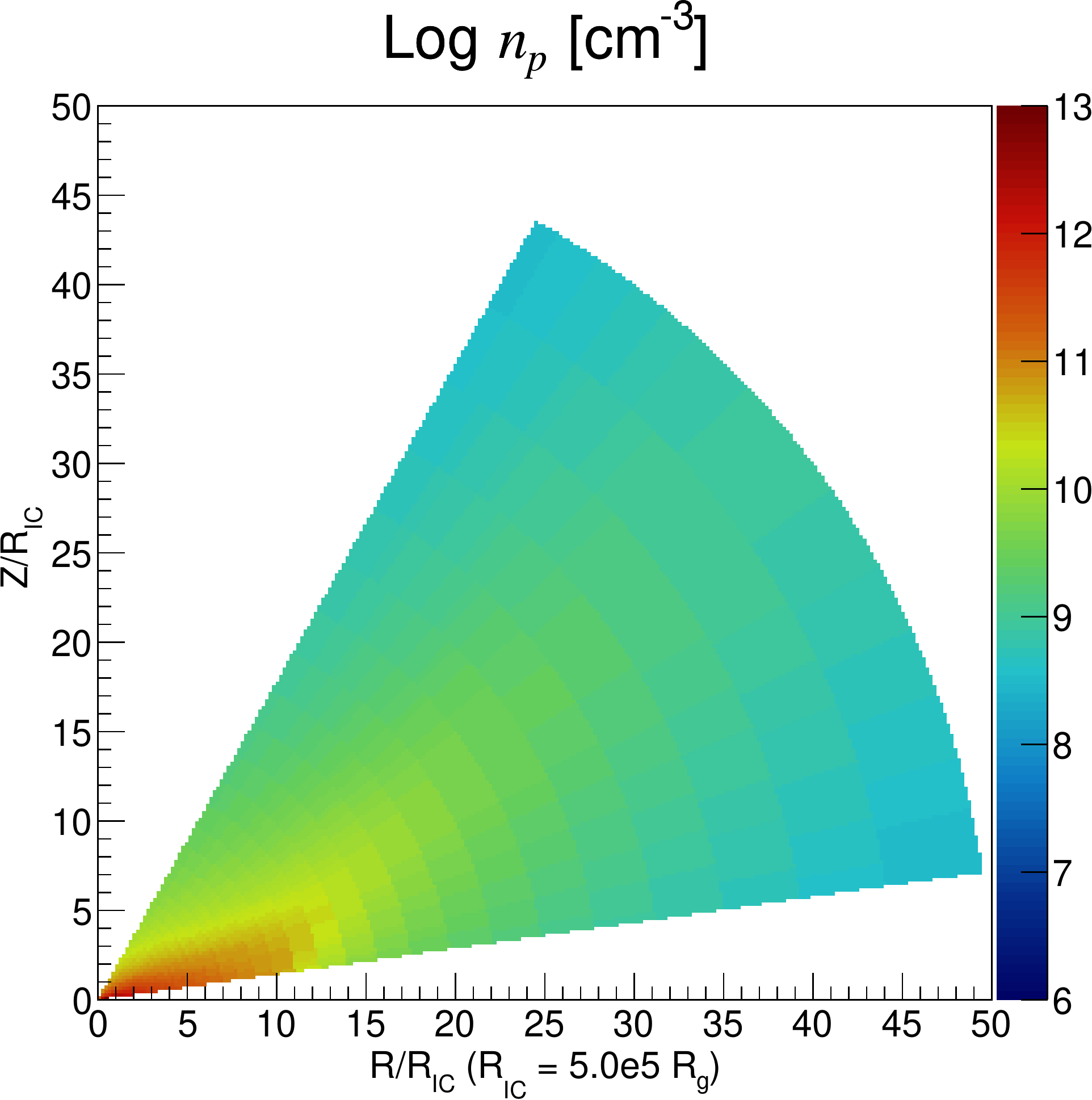}
    \includegraphics[width = 0.45\hsize]{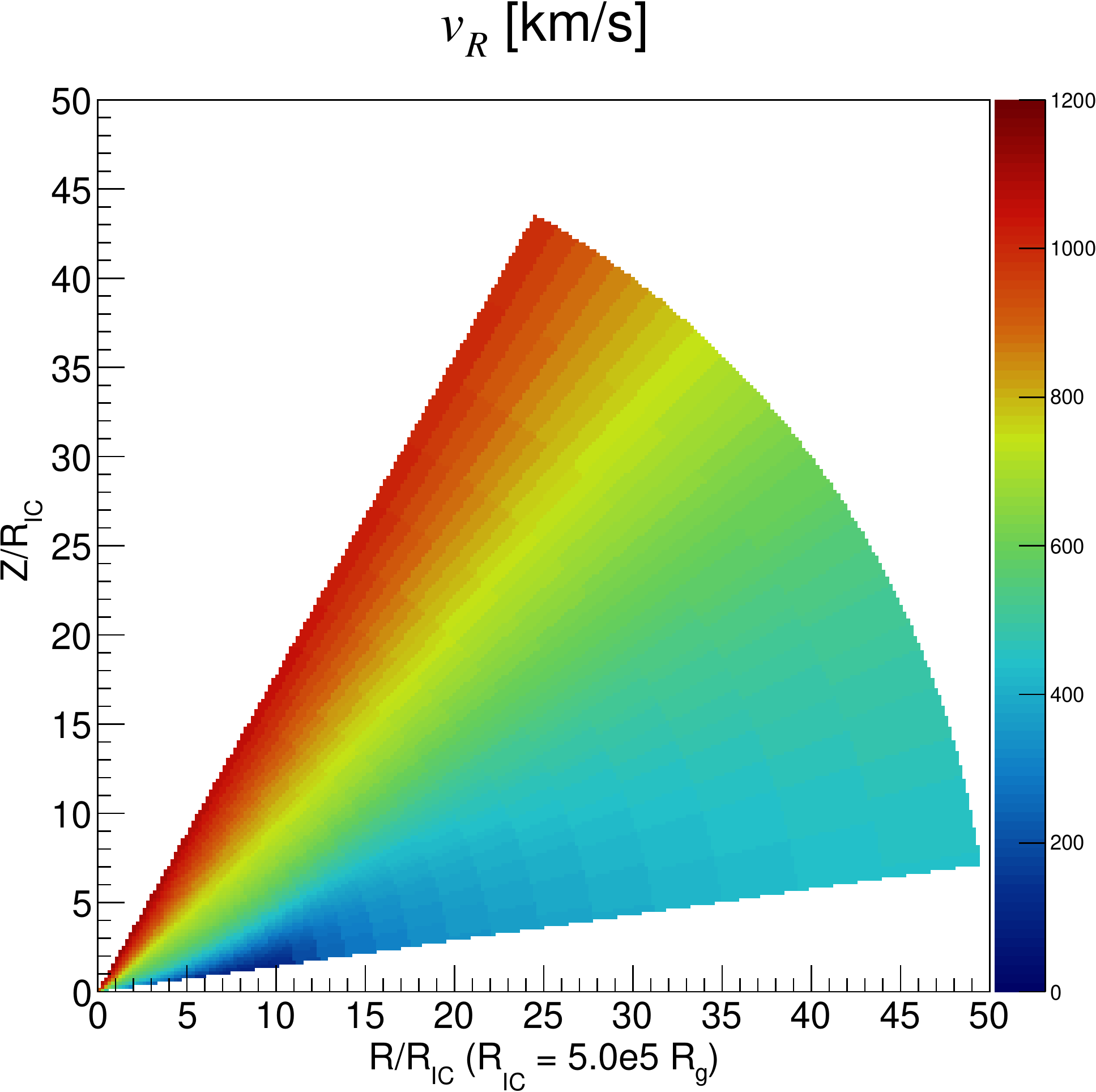}\\
    \includegraphics[width=0.9\hsize]{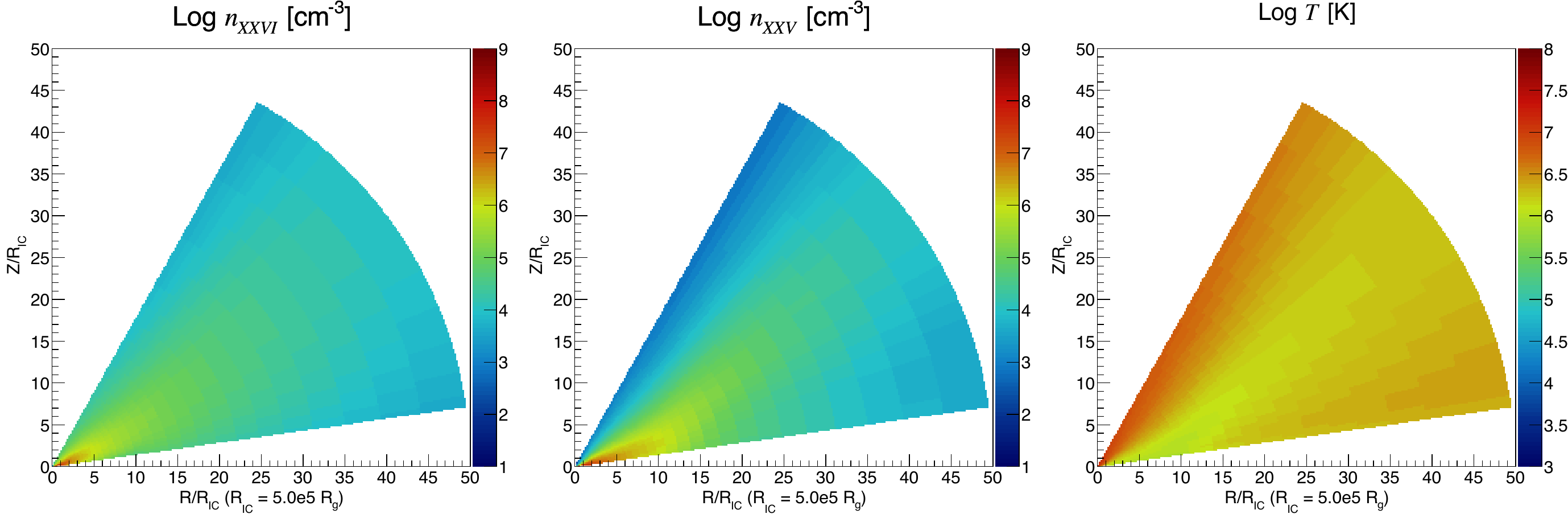}
    \caption{Top: The density and radial   velocity information from the radiation  hydrodynamic code.
    These data are input into {\sc cloudy} to get the detailed ion population for input into the radiation transfer code {\sc monaco}.
    These are shown in the bottom panel for Fe {\sc xxv} and {\sc xxvi}, together with the derived temperature.}
    \label{fig:GX13p1_rebin}
\end{figure*}
\fi

\begin{table}
    \centering
    \caption{Summary of simulations and inner corona parameters for GX 13+1}
    \scalebox{0.725}{
    \begin{tabular}{ccccccc}
    \hline 
    $L/L_\text{Edd}$ & $R_\text{disc}~[R_\text{IC}]$ & $R_\text{ia}~[R_g]$ & $H_c [R_g]$ &$R_\text{is}~[R_\text{IC}]$ & $\dot{M}_\text{w} ~[10^{18}~\mathrm{g/s}]$ & $\dot{M}_\text{w}/\dot{M}_a $\\
    \hline 
    0.5  & 10.0 & 300 & 10 & 0.20  & 9.5 & 8.1 \\
    0.7  & 10.0 & 350 & 13 & 0.22  & 13  & 7.7 \\ 
    \hline 
    \end{tabular}
    }
    \label{tab:summary_GX13p1}
\end{table}

The density and temperature structure resulting from the simulation is shown in Fig.~\ref{fig:GX13p1_hydro}.
The density plot clearly shows how the streamlines splay outwards for $R \gg R_\text{disc}$, 
and this increased adiabatic cooling pulls the gas temperature down below the  Compton temperature at large radii. 

We integrate along different lines of sight to get ion column densities as a function of inclination angle.
We do this for Si {\sc xiv} (magenta) and  S {\sc xvi} (blue) as well as H- and He-like iron (red and green),
and compare these with the observed columns inferred from the fitting above (Fig.\ref{fig:GX13p1_column}). 
The predicted ion columns are well matched to the data,
without any free fitting parameters, showing that the thermal wind simulation gives a very good overall match to the observed properties of the wind in GX 13+1.

\subsection{Changing luminosity: $L/L_\text{Edd} = 0.7$}

The material in the wind is highly ionised, 
so while the ion columns derived above are quite small,
the total column is quite large, of order $10^{24}$~cm$^{-2}$ at a high inclination (Fig.~\ref{fig:NH}).
Electron scattering reduces the observed luminosity along these lines of sight, so the intrinsic luminosity is underestimated. 
Hence we run a new simulation, increasing the luminosity to $0.7~L_\text{Edd}$ and compare the ion column densities and their velocities (dashed-dot lines) to the previous fiducial simulation (solid lines) and the data (dashed points) in Fig.\ref{fig:GX13p1_column}. 
Surprisingly, there is little change in the predicted column densities, and only a slight increase in predicted velocity.
This result is very different to the highly simplified radiation pressure correction of D18 which predicts a much larger column as $L\to 0.71L_\text{Edd}$ due to radiation pressure reducing the effective gravity, allowing the wind to be launched from closer in. 

The difference is due to the effect of the X-ray heated atmosphere over the inner disc.
This effect was not included in D18, but here it effectively limits the launch radius 
of the wind to the radius at which the disc curvature allows it to emerge from the 
shadow, $R_\text{is}\sim 0.2~R_\text{IC}$.
This radius is around the classic thermal wind launch radius, so increasing the luminosity does not decrease the radii at which the wind can form.
In fact, the increased luminosity predicts a slightly increasing scale height of the inner X-ray heated disc atmosphere, so gives a somewhat larger shadowed region.
We strongly caution that the calculation of this inner attenuation zone is sub-grid physics which is not included in our radiation hydrodynamic simulation.
It is instead based on analytic approximations which do not include the effects of radiation pressure (\citealt{Begelman1983b}, T19).
Nonetheless, we physically expect the shadow to be still present, and for its extent not to change too dramatically between $L=0.5\to 0.7L_\text{Edd}$, 
so the inner edge of the wind is fixed at $\sim 0.2R_\text{IC}$ rather than decreasing to encompass most of the inner disc as in D18. 
In this case, the overall mass loss rate increases more or less linearly with the increase in luminosity so that the wind efficiency remains constant,
with $\dot{M}_w\propto \dot{M}_{a}$ (Tab.~\ref{tab:summary_GX13p1}).
Thus the total mass loss increases with the luminosity, but the radiation force increases the wind velocity,
so the total ion columns remain somewhat similar as mass continuity sets the wind density $\propto \dot{M}_w/v_{R}$. 
This result suggests that ion columns are not as sensitive to changing luminosity in the high $L/L_\text{Edd}$ regime as predicted in D18.
Instead, the shadow cast by the inner attenuating zone is crucial in determining the extent of the wind.

\begin{figure}
    \centering
    \includegraphics[width = 0.9\hsize]{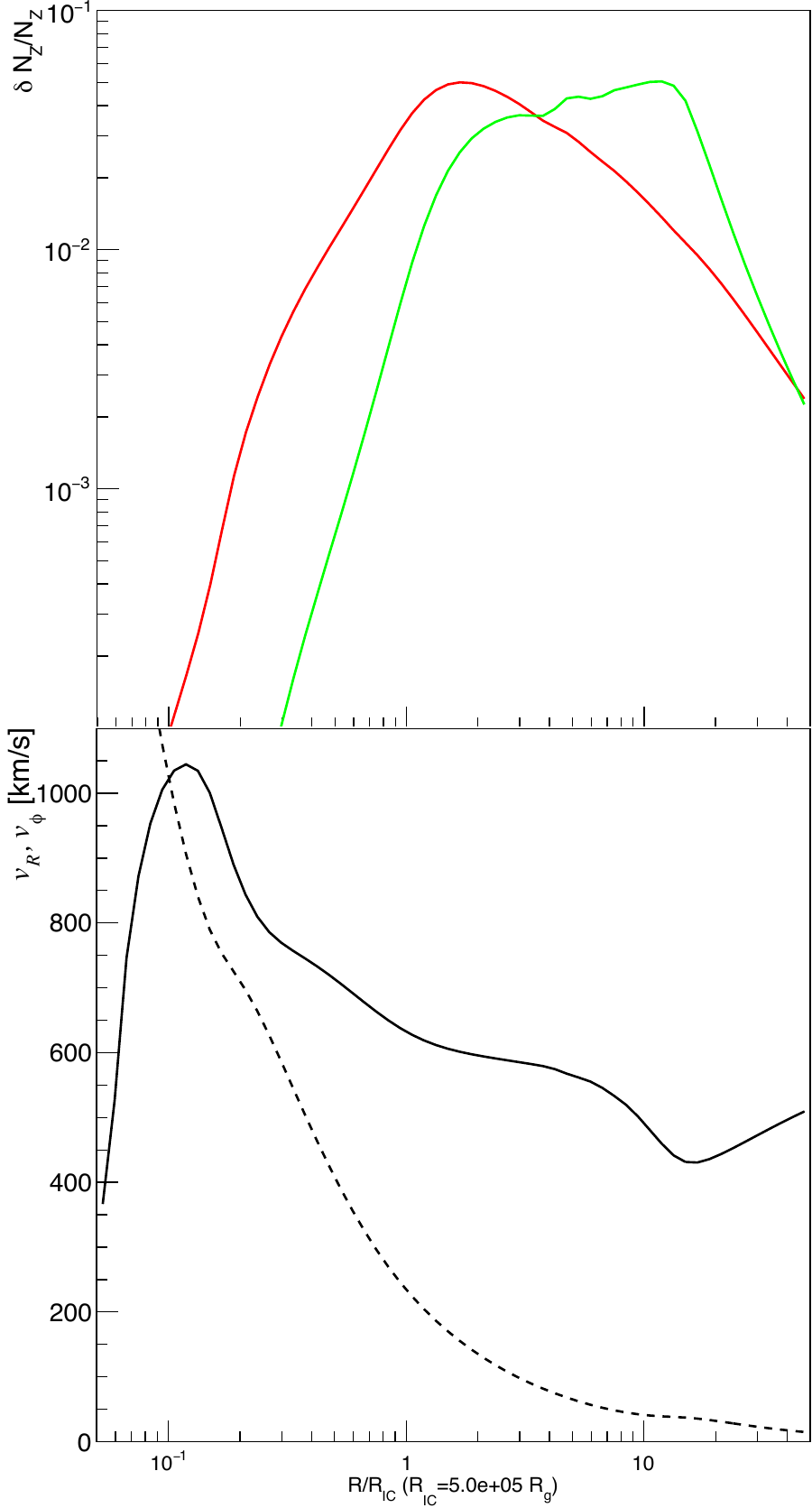}
    \caption{Top: the ion column densities in Fe {\sc xxv} (green) and {\sc xxvi} (red) along a sightline inclined at $65^\circ$. Bottom: Radial velocity (solid) and rotational velocity (dashed) along the same sightline.
    }
    \label{fig:radial_profile_GX13p1}
\end{figure}

\section{Monte Carlo radiation transfer}

\subsection{Calculation setup}

%\begin{figure}
%    \centering
%    \includegraphics[width=0.9\hsize]{figures/difference_hydro_cloudy-crop.pdf}
%    \caption{The difference of ion columns between the result of radiation %hydrodynanmic simulation (solid lines) and that of {\sc cloudy} (dot-dashed lines).
%    Colors show Fe {\sc xxvi} (red) and Fe {\sc xxv} (green).
%    }
%    \label{fig:diff}
%\end{figure}

We calculate the detailed radiation transfer through the hydrodynamic simulations as in T19. We reduce the $\theta$ grid to 
$30-80^\circ$ in order to speed up the calculation, as we now also include an azimuthal grid so that the radiation transfer is fully 3D.
The total grids we use are 60 (radial) and  51 (polar) and 32 (azimuth).
%The input density and velocity are shown in Fig.\ref{fig:GX13p1_rebin}.

Those reduced grid's density and velocity are input 
into the radiation transfer code {\sc monaco} \citep{Odaka2011}.
This code uses the Geant4 toolkit library \citep{Agostinelli2003} for photon tracking in arbitrary three-dimensional geometry,
but has its modules handling photon interactions \citet{Watanabe2006,Odaka2011} 
so that it can treat the interactions  such as photo-ionisation or photo-excitation,  and photons generated via recombination and atomic de-excitation.
The energies and oscillator strengths for the H and He-like ions were calculated from the Flexible Atomic Code \citep{Gu2008} as detailed in  Tab.~\ref{tab:line id}. 

{\sc monaco} also handles electron scattering and the Doppler shift of the absorption cross-section from the velocity structure of the material. 
This cross-section is calculated for the photon energy in the comoving frame and Lorentz transformed back into the rest frame.
The Doppler broadening of temperature and turbulent motion is also considered.
The number of input photons is $1.4\times 10^8$ between 6.5-8.5~keV which is the same number as in T20.

{\sc monaco} requires the distribution of ion populations and temperature in addition to density and velocity.
We obtain these more accurately than is possible in the more approximate approach of the 
radiation hydrodynamic code by solving the one dimensional radiation transfer along line of sight using {\sc cloudy}.
We chain {\sc cloudy} radially through the density structure and use the output spectrum of the inner grid as the input spectrum to the next grid.
This calculation gives the resulting ion columns of Fe {\sc xxv} and {\sc xxvi} for input into {\sc monaco}, along with the self-consistent temperature. We check that these do not actually differ much from the original results of the radiation hydrodynamic code.
%(bottom panels of Fig.\ref{fig:GX13p1_rebin}).

The upper panel of Fig. \ref{fig:radial_profile_GX13p1} shows the radial distribution of these ion columns for a representative line of sight at 65$^\circ$. This shows that the grid is large enough to resolve where the ion columns peak. The
lower panel shows the corresponding 
radial (solid) and azimuthal (dashed) velocity (lower panel) of the material. 
The peak radius at which the higher ionisation Fe {\sc xxvi} ($\sim R_\text{IC}$) is produced is smaller than that of Fe {\sc xxv} ($\sim 10 R_\text{IC})$. 
The corresponding radial velocity is $\sim 600~\mathrm{km/s}$ (Fe {\scriptsize XXVI}) and $\sim 400~\mathrm{km/s}$ (Fe {\scriptsize XXV}), while the rotational velocity is lower, at  $\sim 200 \text{km/s}$ (Fe {\scriptsize XXVI}) and $\sim 50 \text{km/s}$ (Fe {\scriptsize XXV}). 

\begin{figure*}
    \centering
    \includegraphics[width=0.9\hsize]{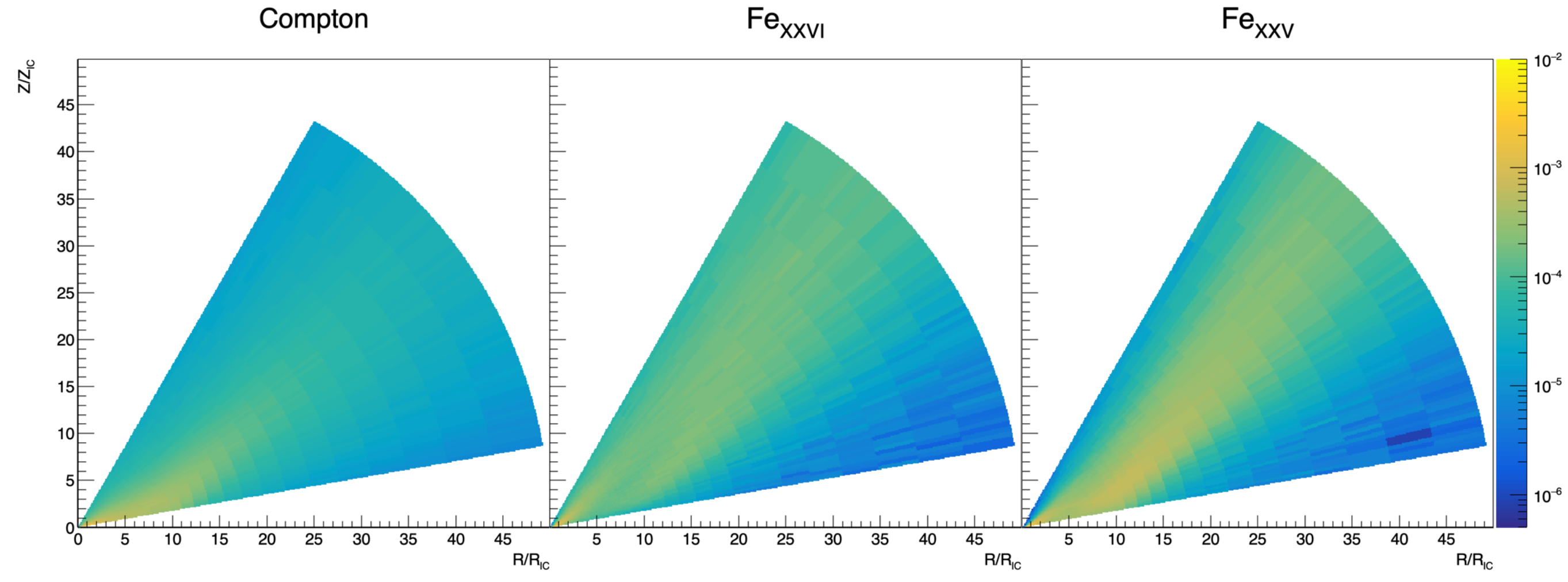}
    \caption{The distributions of the probability for the Compton scattering (light), photo-excitation of Fe {\sc xxvi} (middle), and that of Fe {\sc xxv} (right). }
    \label{fig:prob pos}
\end{figure*}

Fig.\ref{fig:prob pos} shows the 2D distribution of total column (a), Fe  {\sc xxvi} (b) and {\sc xxv} (c), colour coded as a probability for scattering/emission. Again this shows that the grid can resolve most of the ion emission region, 
especially the highest ionisation ion, Fe  {\sc xxvi}. 
Completely ionised material peaks at even smaller radii, so the Compton scattered flux peaks at even smaller radii than that of Fe {\sc xxv} because the inner region has a higher ionisation state.

\begin{figure*}
    \centering
    \includegraphics[width=0.9\hsize]{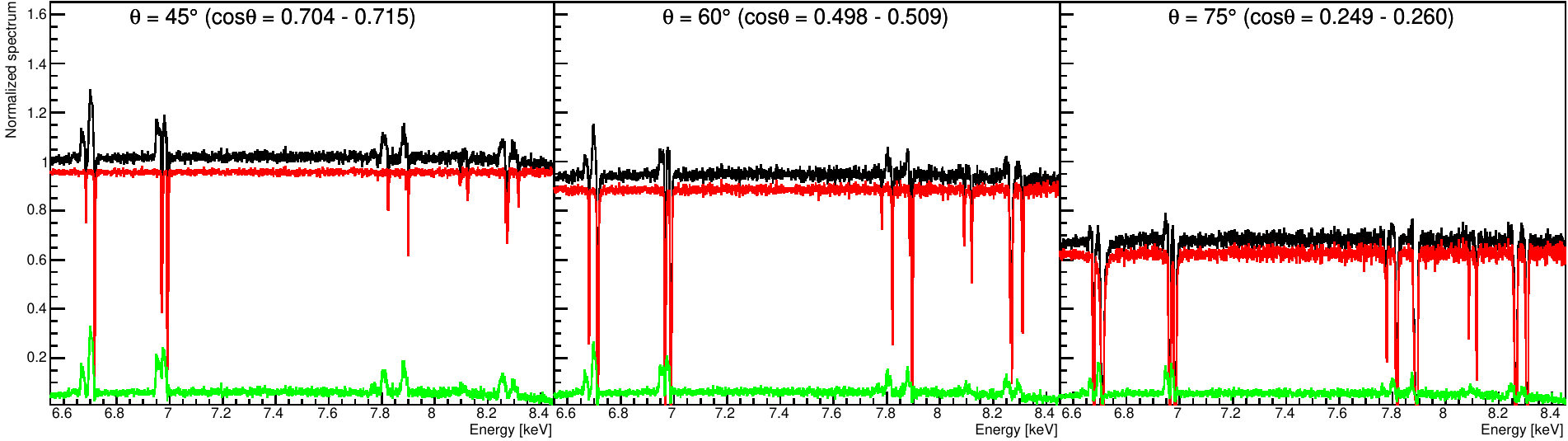}
    \caption{The predicted spectral features in the iron line region at three inclination angles for 
    $v_\text{turb} = 0$. Red shows the transmitted spectrum, green shows the emission/scattered flux in the wind, and black shows the total (observed) spectra. 
    %$v_\text{turb} = 0$ (top), $ v_{R}$ (middle), and $v_\phi$ (bottom). Red shows the transmitted spectrum, green shows the emission/scattered flux in the wind, and black shows the total (observed) spectra. 
    }
    \label{fig:GX13p1_angle}
\end{figure*}

\begin{figure*}
    \centering
    \includegraphics[width=0.3\hsize]{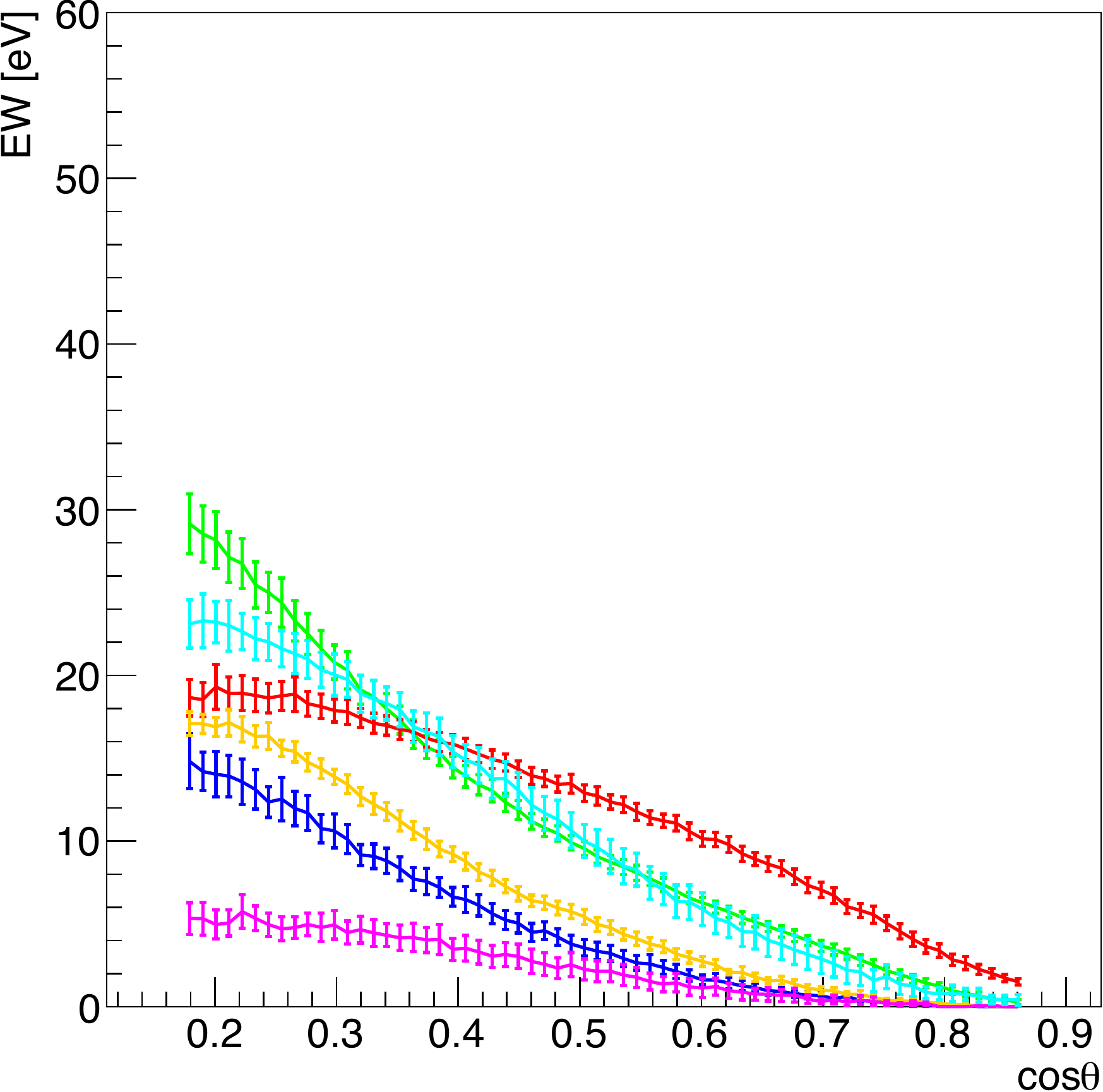}
    \includegraphics[width=0.3\hsize]{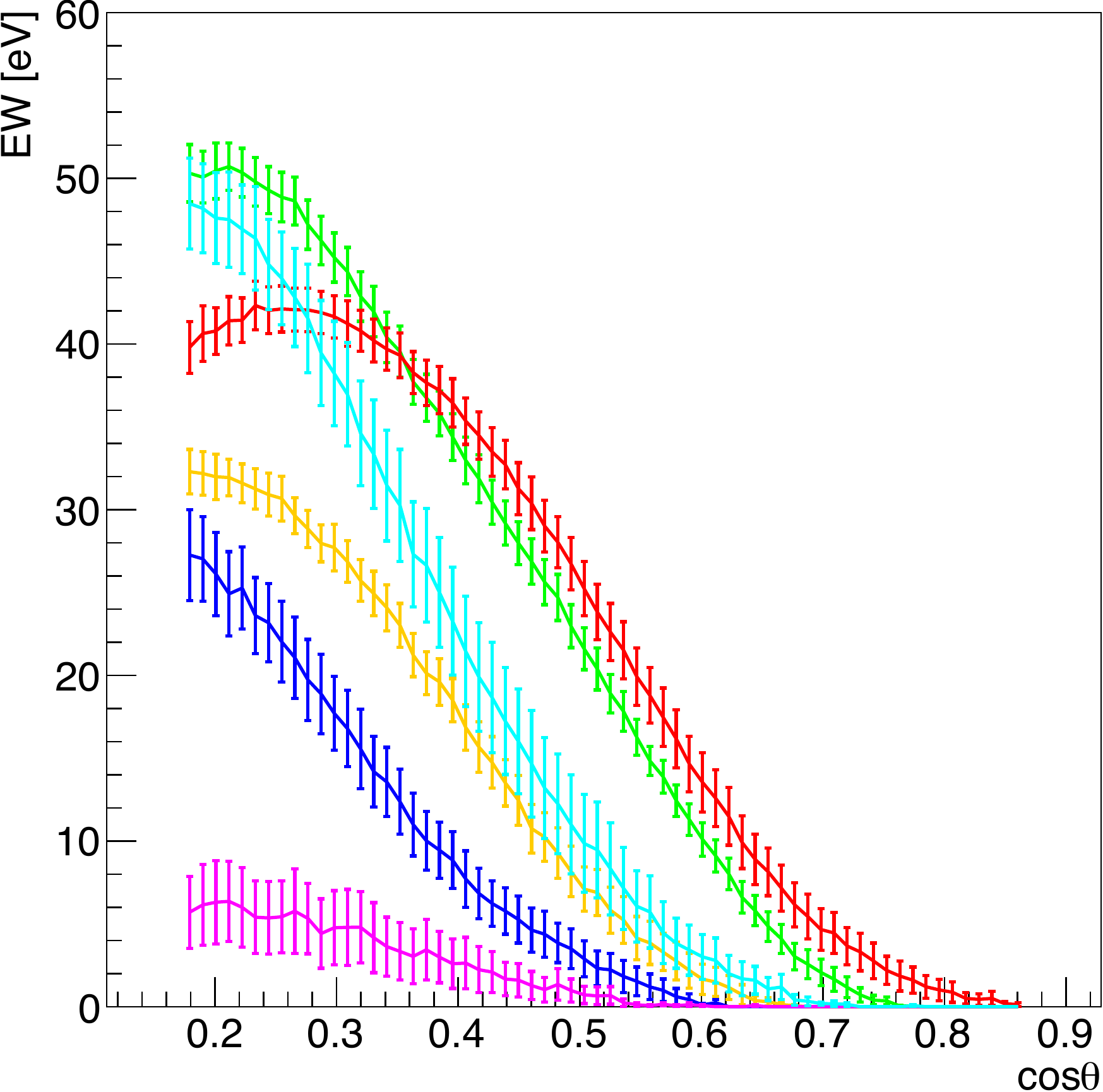}
    \includegraphics[width=0.3\hsize]{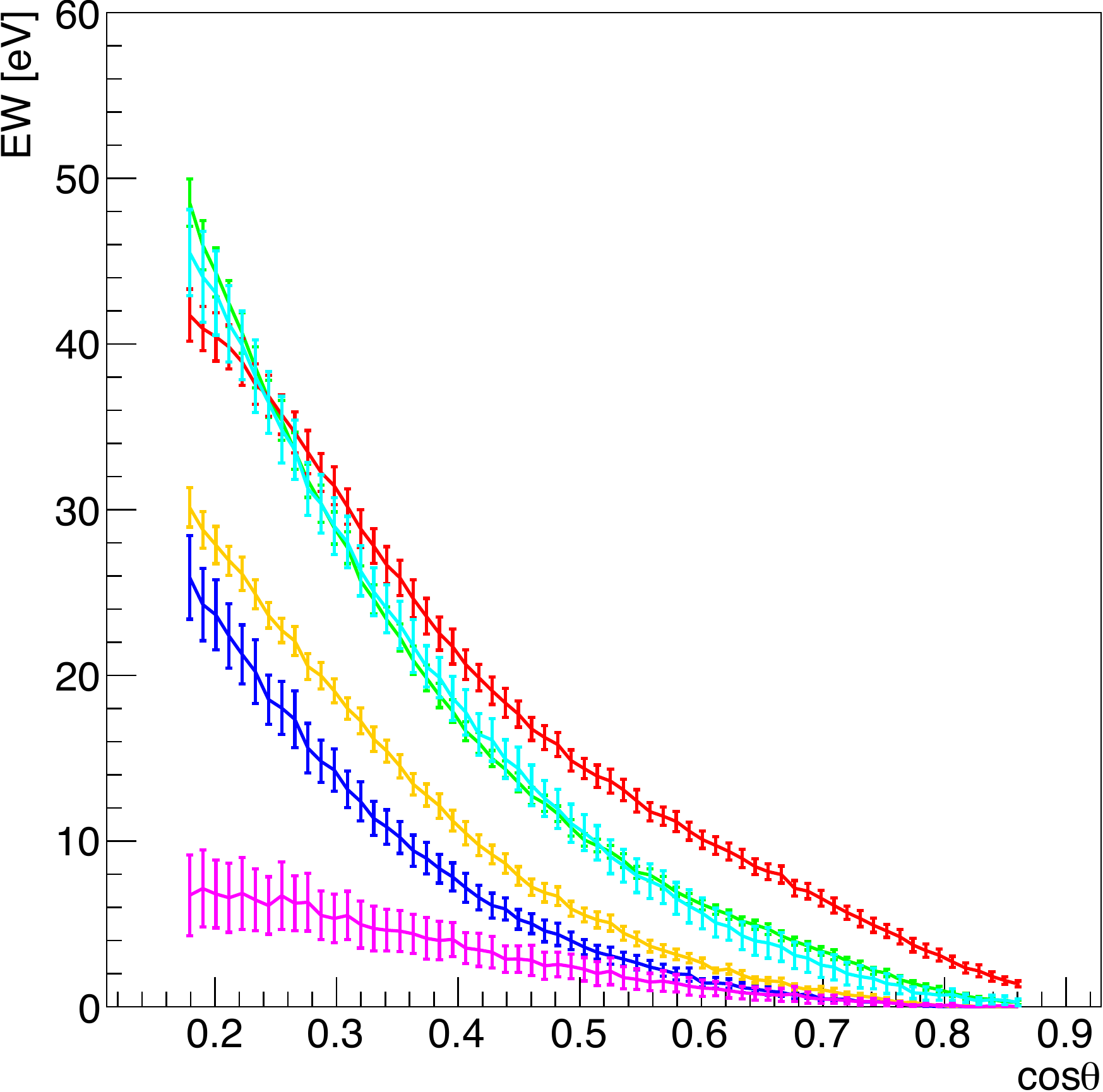}
    \caption{The inclination dependence of the absorption line EWs for $v_\text{turb}= 0$ (left), $v_{R}$ (middle) and $v_\phi$ (right)
    Colors show Fe {\scriptsize XXV} He$\alpha~y + w$  (green),
    Fe {\scriptsize XXVI} Ly$\alpha_1+\alpha_2$ (red),
    Ni {\scriptsize XXVII} He$\alpha~y+w$ (blue), 
    Fe {\scriptsize XXV} He$\beta~w$ (orange), 
    Ni {\scriptsize XXVIII} Ly$\alpha_1+\alpha_2$ (magenta), and 
    Fe {\scriptsize XXVI} Ly$\beta_1+\beta_2$ + Fe {\scriptsize XXV} He$\gamma~w$ (cyan). The effect of saturation can clearly be seen as the line EWs increase with increasing turbulent velocity.
    Note that error bars are calculated by Poisson noise.
    }
    \label{fig:EW_GX13p1}
\end{figure*}

\begin{figure*}
    \centering
    \includegraphics[width=0.9\hsize]{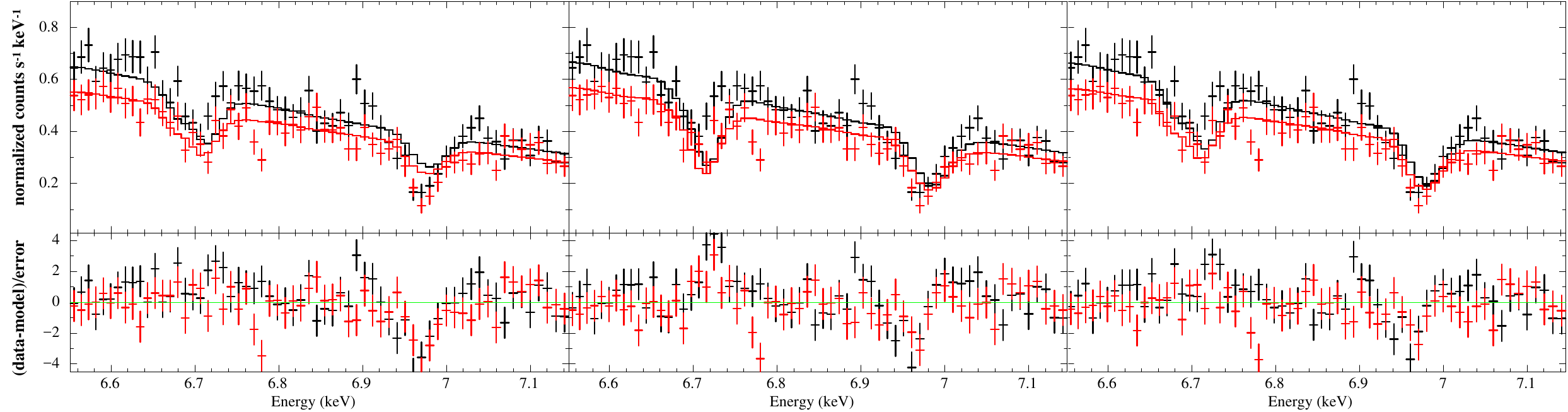}
    \caption{The observed Chandra first order HEG spectra (OBSID: 11818) of +1 (black) and -1 (red) compared to our models with  
     $v_\text{turb} = 0 $(left), $v_R$ (middle), and $v_\phi$ (right).
     The best fit inclination angles for each are 
     $\cos\theta = 0.29, 0.47, 0.33$
     }
    \label{fig:GX13p1_1st}
\end{figure*}

\begin{figure*}
    \centering
    \includegraphics[width=0.9\hsize]{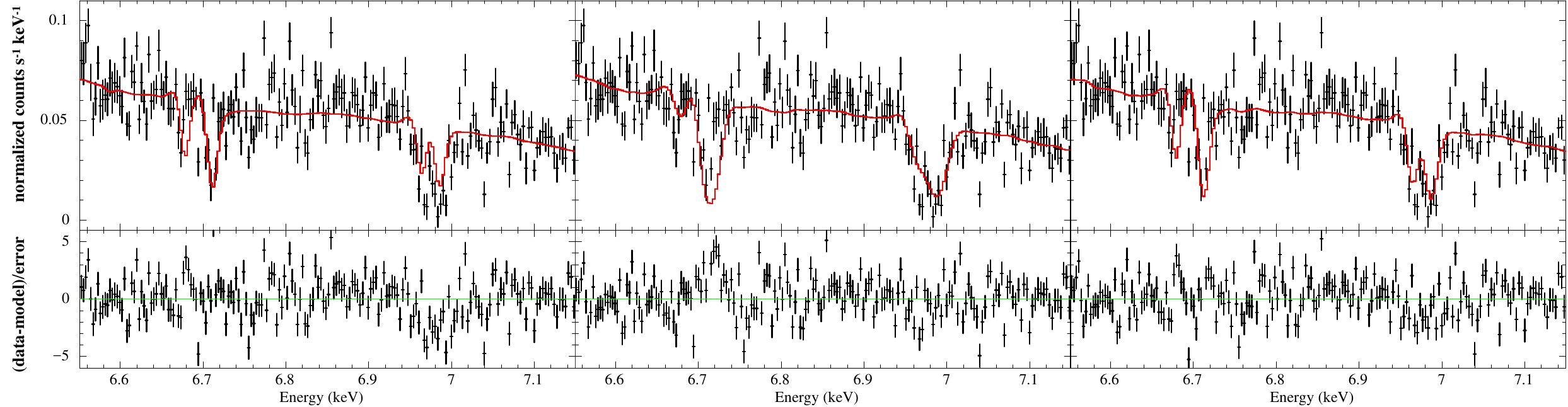}
    \caption{The observed Chandra third order HEG spectra (co-added from the 4 TE observations) compared to our models with
     $v_\text{turb} = 0 $(left), $v_R$ (middle), and $v_\phi$ (right). It is clear that the models with $v_\text{turb} =v_R$ overestimate the observed line widths. All models underestimate the depth of the Fe {\sc xxvi} K$\alpha_{1,2}$ lines, and overestimate the blueshift of the Fe {\sc xxv} $y$ intercombination line. 
     The best fit inclination angles for each are 
     $\cos\theta = 0.35, 0.47, 0.37$. 
     }

    \label{fig:GX13p1_3rd}
\end{figure*}

\subsection{The effect of additional turbulent velocity}

%Because of the large EWs, column density of Fe {\scriptsize XXVI} and {\scriptsize XXV} has large uncertainty.

The ion columns are large, so the line profiles are 
strongly dependent on the turbulent velocity (see also T20).
%We run {\sc monaco} with $v_\text{turb} = 0 $ and  $v_\text{turb}, v_\text{R}$. 
Fully developed turbulence has $v_\text{turb}\sim v_\text{flow}$.
However, there are two potential flow velocities in our simulation as we now have a fully 3D structure.
One is the radial velocity, $v_\text{R}$, and the other is the azimuthal velocity $v_\phi $ (lower panel of Fig.~\ref{fig:radial_profile_GX13p1}). 
Our 2D radiation hydrodynamic simulation can capture some aspects of $R-\theta$ turbulence, so the lack of vortices in our results may signal that this is not present (see also \citealt{Woods1996}).
However, there is also an azimuthal shear layer between the disc atmosphere and the wind which is not included in our 2D simulation,
and which could give rise to turbulence via the Kelvin-Helmholtz instability. 
Thus,  we explore the effect of turbulence by running {\sc monaco} with $v_\text{turb} = v_\phi $ in addition to $v_\text{turb} = 0$ and $v_\text{turb} = v_{R}$.
%Detailed fits to the {\it Chandra}/HETGS line profiles already show that the widths are less than $\sim 200$~km/s \cite{Allen2018},
%so $v_{R}\sim 400-600$~km/s is likely too large, but we include it here in order to show this explicitly. 

We plot the resulting spectra with $v_\text{turb} = 0 $ at three different angles in Fig. \ref{fig:GX13p1_angle} (The bin size is 1 eV).
%The shape of spectra is plotted in Fig. \ref{fig:GX13p1_angle}. 
The behaviour of these profiles show a similar trend to that in H1743-322 (T20), i.e. the absorption lines are stronger at higher inclination angle, as is 
attenuation due to electron scattering in the wind.  
By contrast, the scattered/emitted flux from the wind (the green line, which is $\sim 0.05$ of the intrinsic continuum level) is constant with inclination. 
However, unlike H1743-322, the Fe {\sc xxv} and {\sc xxvi} K$\alpha$ absorption lines often go to zero intensity in the transmitted spectrum  (red line), 
i.e. are saturated due to the higher column density in GX13+1 due to its larger disc.
However, the contribution of the scattered flux means that the absorption lines in the total spectrum (black) never go entirely to zero at their core. 

We explore the effect of turbulence as well as inclination angle on the saturation by 
quantifing the predicted absorption line EWs of all the lines in the
6.5-8.5~keV bandpass as in T18 (Fig. \ref{fig:EW_GX13p1}), i.e.
by fitting the line-free regions with a quadratic to estimate the continuum level and by computing EWs numerically. 
The Ni {\sc xxviii} (H-like) K$\alpha$ lines (magenta) are the only lines with identical EW in all simulations.
All the other lines increase in EW with increasing turbulence, so are 
saturated to a greater or lesser 
extent. Even the Helium-like 
Ni K$\alpha$ $w$ (resonance) line goes optically thick at high inclinations.
The only unsaturated lines are the 
K$\alpha$ $y$ (intercombination) line of Ni {\sc xxvii} at 7.765~keV, along with the H-like Ni {\sc xxviii} K$\alpha$ resonance lines at 8.073 and 8.102~keV (as expected from the oscillator strengths and abundances in Tab~\ref{tab:line id}).

Unfortunately, due to its small effective area, spectra taken from  {\it Chandra}/HETGS are not sensitive to the more unsaturated transitions above 7~keV (see Tab.~\ref{tab:line id}).
However, observations using {\it XMM-Newton} have reported EWs of these lines  \citep{DiazTrigo2014}
of $\sim 20~\text{eV}$ around 8.2 keV coming from Fe {\scriptsize XXVI} Ly$\beta_{1, 2}$+ Fe {\scriptsize XXV} He$\gamma~w$, and $\sim 15~\text{eV}$ around 7.9 keV coming from Ni {\scriptsize XXVII} He$\alpha~w+y$ + Fe {\scriptsize XXV} He$\beta~w$.
Those observed EWs in high energy band have large uncertainty but are consistent with our estimates here only for the simulation without additional turbulence. 
Future observations with the calorimeters on {\it XRISM} and {\it Athena} will show this region of the spectrum with much higher accuracy,
and tightly constrain the level of saturation. 

\subsection{Fitting to {\it Chandra}/HETGS first order spectra}

\if0
\begin{table}
    \centering
    \caption{Fitting results}
    \begin{tabular}{ccccc}
    \hline
    OBSID & $\chi^2/\nu~ (v_\text{turb} = 0) $ & $\chi^2/\nu~(v_\text{turb}= v_{R})$ & $\chi^2/\nu~(v_\text{turb} = v_\phi)$ &$\chi^2/\nu~(R_\text{disc} = R_\text{IC},~ v_\text{turb} = v_\phi)$\\
    \hline
    11815 & 429/124 & 192/124 & 168/124 & 152/124\\
    11816 & 345/124 & 250/124 & 223/124 & 200/124\\
    11814 & 174/124 & 180/124 & 128/124 &123/124\\ 
    11817 & 262/124 & 259/124 & 158/124 & 151/124\\
    11818 & 215/124 & 225/124 & 171/124 & 161/124 \\
    \hline
    \end{tabular}
    \label{tab:fit_result_GX13p1}
\end{table}
\fi 
\begin{table}
    \centering
    \caption{Fitting results to all {\it Chandra}/HEG first order spectra}
     \scalebox{0.9}{
    \begin{tabular}{cccc}
    \hline
    OBSID & $\chi^2/\nu~ (v_\text{turb} = 0) $ & $\chi^2/\nu~(v_\text{turb}= v_{R})$ & $\chi^2/\nu~(v_\text{turb} = v_\phi)$\\
    \hline
    11815 & 429/124 & 192/124 & 168/124 \\
    11816 & 345/124 & 250/124 & 223/124 \\
    11814 & 174/124 & 180/124 & 128/124 \\
    11817 & 262/124 & 259/124 & 158/124 \\
    11818 & 215/124 & 225/124 & 171/124  \\
    \hline
    \end{tabular}
    }
    \label{tab:fit_result_GX13p1}
\end{table}
\begin{figure}
    \centering
    \includegraphics[width=0.9\hsize]{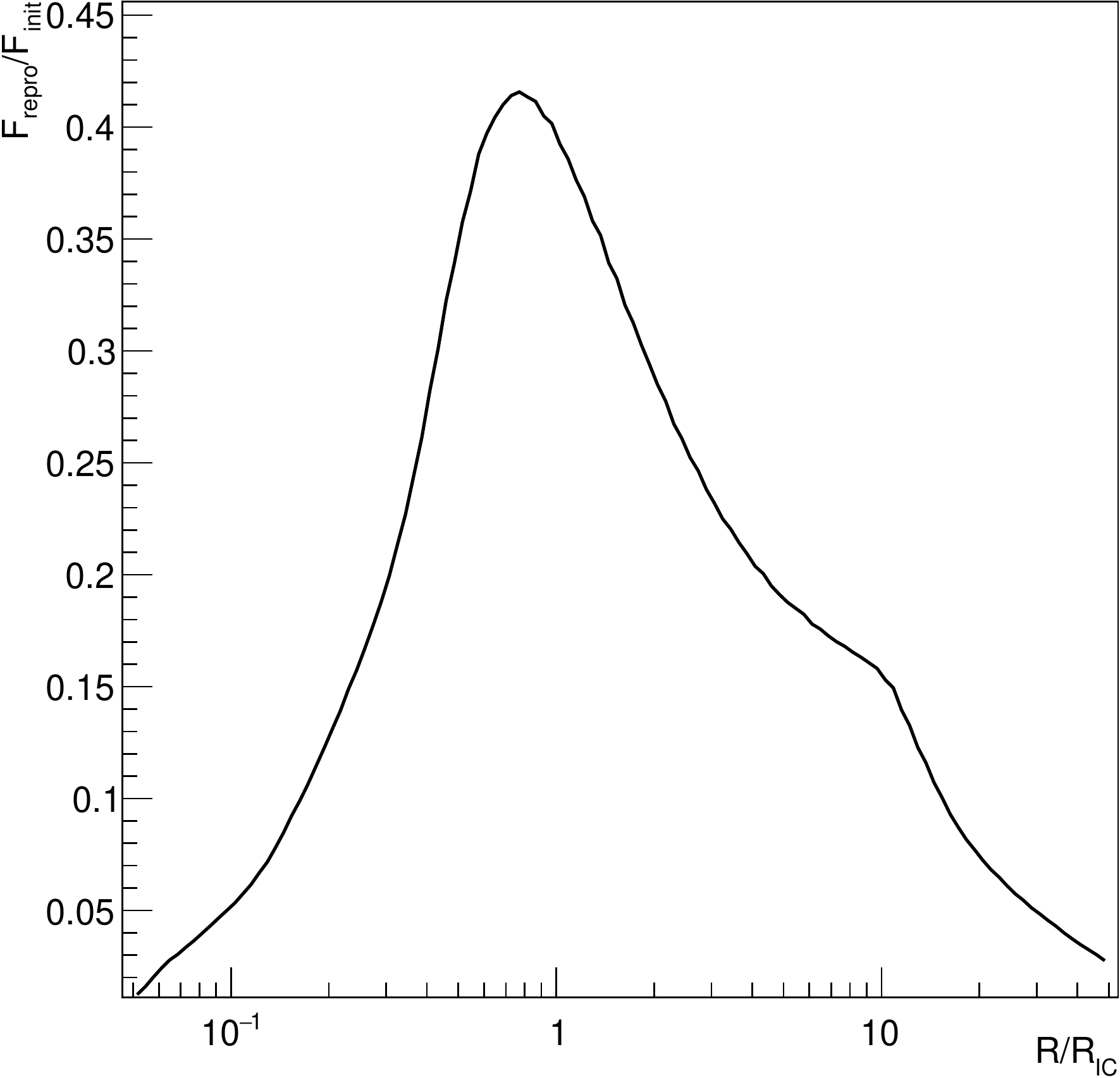}
    \caption{The radial distribution of the ratio scattered flux to direct flux along a sightline at $70^\circ$. }
    \label{fig:scattered}
\end{figure}
We now directly fit these models to the data taken from $\pm1$ orders of the High Energy Grating (HEG) (ObsID:11818 CC mode) (Fig.\ref{fig:GX13p1_1st}). 
Colors in Fig.\ref{fig:GX13p1_1st} show +1 (black) and -1 (red) in HEG and two lines are the same models but different response of $\pm1$ orders.
Fig.~\ref{fig:GX13p1_1st} shows that all the different assumptions on the turbulence leave some residuals,
but the fit quality is better for models with  $v_\text{turb} = v_\phi$.
This trend is also seen in fits to the TE mode spectra 
(Tab.\ref{tab:obs_list_GX13p1}).% though these are more affected by pileup (see \citealt{Allen2018}). 
These all show consistently that the models with $v_\text{turb}=v_\phi$ have the lowest $\chi^2$, then those with $v_\text{turb}=0$, and finally $v_\text{turb}=v_R$.%, as expected from the line velocity widths in \cite{Allen2018}.
%\textcolor{red}{EWs of Fe {\sc xxvi} in this sources are 30-50 eV \citep{Allen2018},  therefore 
%those of the model with $v_\text{turb} = 0$ is  smaller than those values (red in the left panel of Fig.\ref{fig:EW_GX13p1}). 
%This is because $\chi^2$ of this model is worse than that of $v_\text{turb}= 0$ even that the column of Fe {\sc xxvi} obtained from our radiation hydrodynamic simulation is consistent with observation.}

We can now look in more detail at the line profiles using third order HEG data from the 4 co-added TE spectra, as this has 
3 times higher resolution than first order \cite{Miller2016}.
Fig.~\ref{fig:GX13p1_3rd} shows the data and our models.
This figure clearly shows that the model with $v_\text{turb}=v_R$ overestimates the observed line widths, but more unexpectedly reveals that the model with $v_\text{turb}=0$ is a good description of the observed line widths. 
There is enough change in radial velocity through the wind to produce the observed line profiles without additional turbulence (see \ref{fig:radial_profile_GX13p1}).
The main issue with the zero turbulence fit is that it underestimates the {\it depth} of the Fe {\sc xxvi} K$\alpha$ doublet compared with that of Fe {\sc xxv}.
This is initially surprising as the ion columns in the model match to those determined by the {\sc kabs} fits (see Fig. \ref{fig:GX13p1_column}).

We first assess whether this is due to the inclusion of emission and scattering in the wind, filling in the absorption line.
However, using only the transmitted spectrum to fit the data shows the same overall mismatch in absorption line strength.
Our model underestimates the EW of Fe {\sc xxvi} K$\alpha_{1,2}$, even with the addition of moderate turbulence.
This is because the simulation is more constrained than single {\sc kabs} fits each absorption line.
The model hardwires the ion column, ionisation state, and velocity structure along each line of sight so has much less freedom than the free ion fits. 
The total ion columns prefer higher inclination angles, while the ratio of Fe {\sc xxvi} to Fe{\sc xxv} prefers lower inclination. Since dips are rare in this source, 
it seems most likely that true inclination is less than the best fit third order HEG fits of $\sim 70^\circ$, so this means that the model slightly underestimates the ionisation state of the wind.

This could be due to the additional illumination from the scattered flux which is underestimated in the 1-dimensional {\sc cloudy} calculation.
We explore this quantatatively using the {\sc monaco} simulation.
Fig. \ref{fig:scattered} shows the ratio of direct to scattered flux as a function of radius along a $70^\circ$ line of sight.
The maximum is at 
most 40  \% at $R\sim R_\text{IC}$ where column of Fe {\sc xxvi} is maximum (Fig. \ref{fig:radial_profile_GX13p1}). 
Hence the ratio between columns of Fe {\sc xxvi} and Fe {\sc xxv}, which is proportional to the ionisation parameter, can be up to 1.4 times larger, and this could at least partly explain the lower ionisation of our model.
However, a more important effect could also be the scattered flux increasing the illumination of the disc surface, which will increase the wind density, which will also increase the scattered flux.
We will explore the non-linear outcome of this in subsequent work.
Other significant uncertainties come from our assumptions about the shadow cast by the inner heated disc atmosphere, which is estimated using an analytic formalism without including the effect of radiation pressure.
We caution that small changes in the subgrid physics of this shadow zone will also affect the detailed structure of the wind,
both the total column density of the material and its ionisation structure.

\if0
One further mismatch between the data and models is the consistent offset in the Fe {\sc xxv} K$\alpha$ $y$ intercombination line.
Our models predict the centroid of this (wind velocity shifted) absorption at 
6.678~keV which is 5eV higher than observed. 
This appears clearly in the intercombination line, but not in the resonance line of the same ion, but this apparent difference is due to their different strengths. The resonance line is also consistent with being shifted back by 5eV, so it seems most likely that this is a velocity shift, with models slightly overestimating the velocity of the Fe {\sc xxv} ion.
We again rule out this being due to the impact of the emission line filling in the red side of the absorption line by fitting the transmitted only spectrum (red line) to the data.
%The same residuals remain. 
\fi 

\section{Predicted {\it XRISM} observation} 
\begin{figure*}
    \centering
    \includegraphics[width=0.45\hsize]{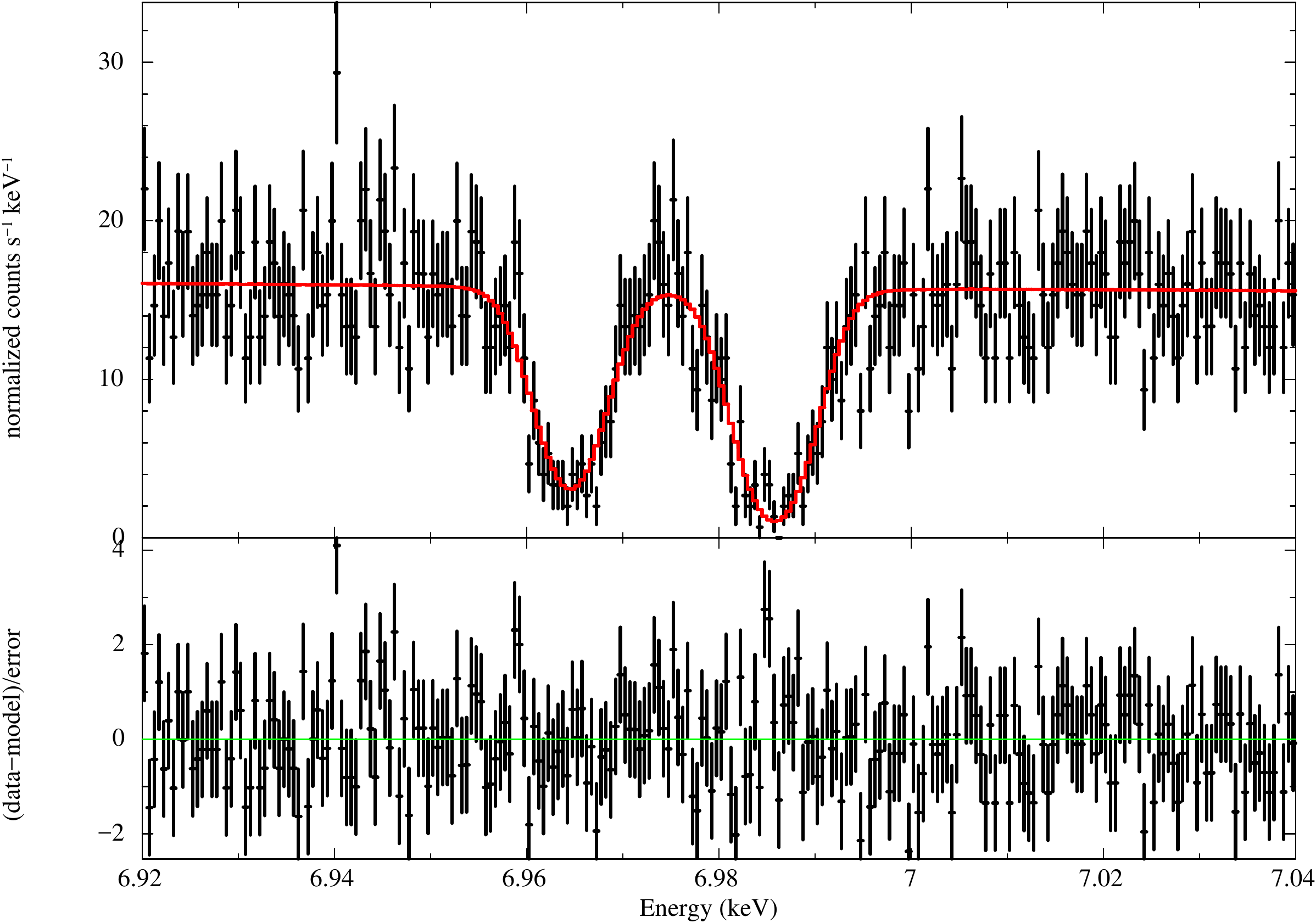}
    \includegraphics[width=0.45\hsize]{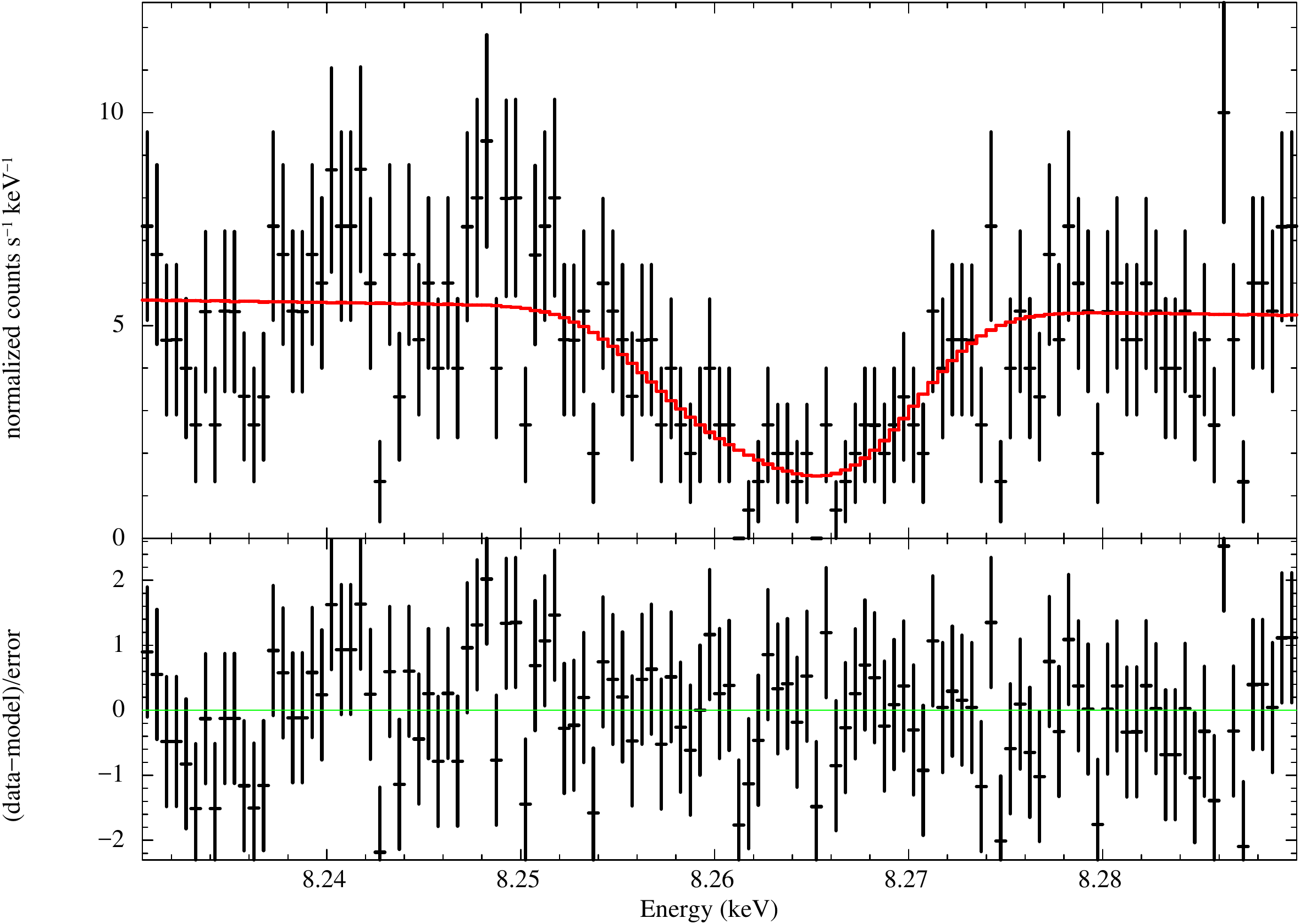} 
    \caption{The predicted absorption lines from Fe {\sc xxvi} Ly$\alpha$ 1,2 (left) and Ly$\beta$ 1,2 (right) with $v_\text{turb} =0$. The red colors show model of {\sc kabs}. }
    \label{fig:Resolve_zero}
\end{figure*}
\begin{figure*}
    \includegraphics[width=0.45\hsize]{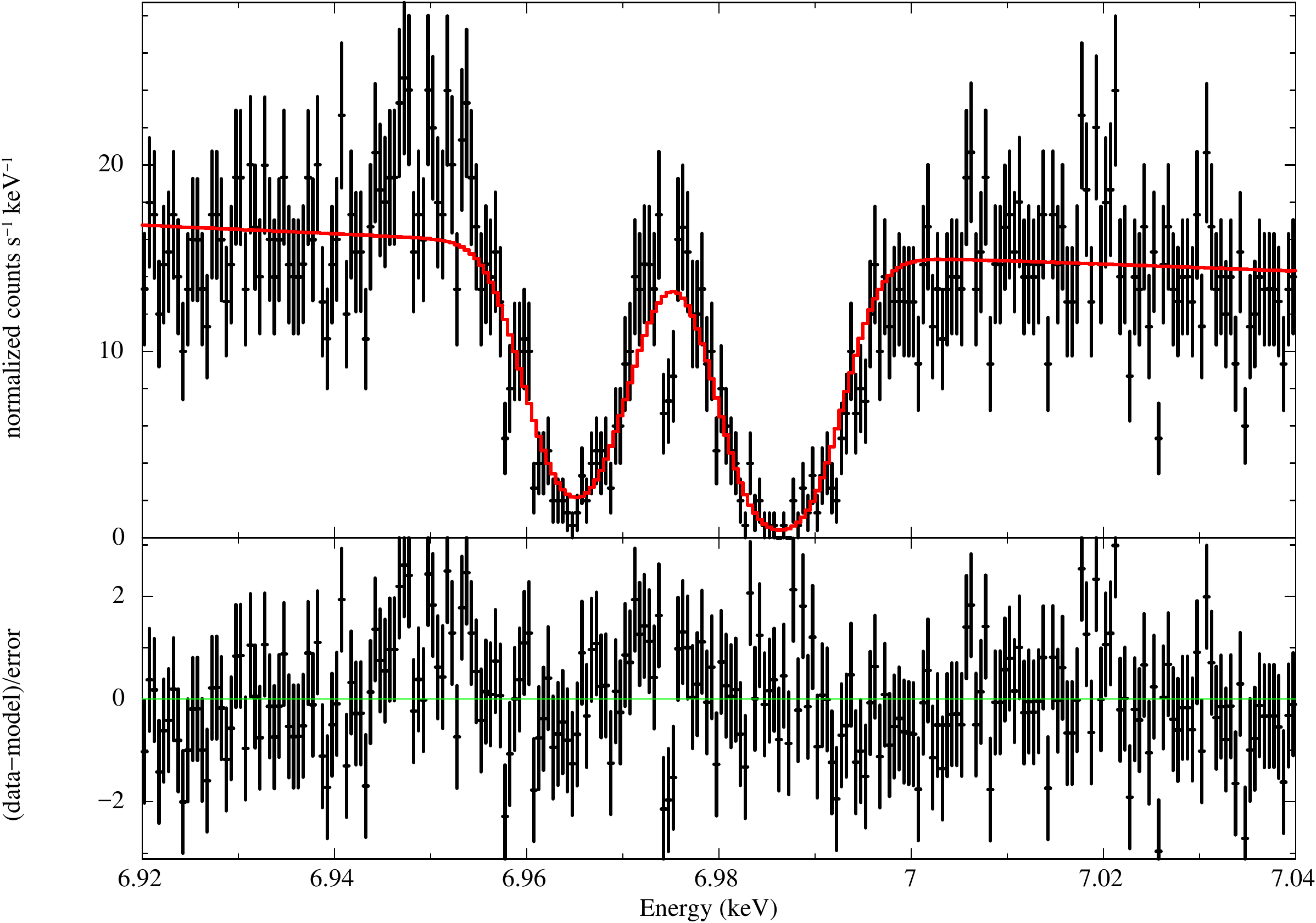}
    \includegraphics[width=0.45\hsize]{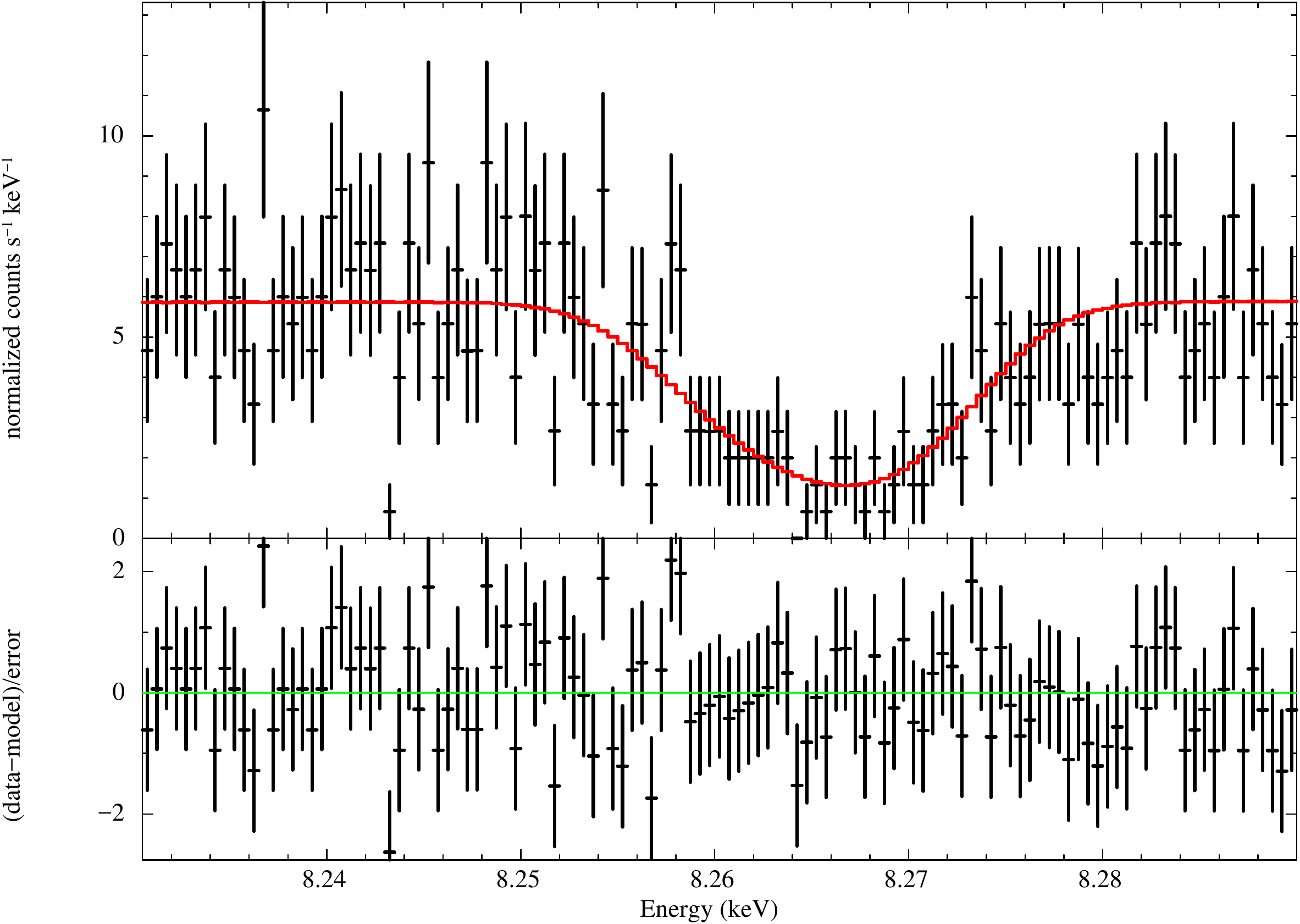}\\
    \caption{As in Fig.\ref{fig:Resolve_zero} but $v_\text{turb}=v_\phi$. 
    The red colors show model of {\sc kabs}. }
    \label{fig:Resolve_phi}
\end{figure*}

We simulate a {\it XRISM}/Resolve observation from the simulation models using the same parameter HEG third order with $v_\text{turb} = 0$ and $v_\phi$ .
The Resolve calorimeter gives 5~eV energy resolution at 6~keV in its best modes.
However, this energy resolution is limited to a count rate of 200 c/s, which corresponds to around 0.1 Crab for an on-axis point source, and this source is brighter than this limit (0.3--0.4 Crab). The count rate can be reduced 
by a factor 10 using the neutral density filter with offset pointing (See T20).
Thus we simulate the 3 ks observation instead of the 30 ks observation, which is the same observation exposure time of {\it Chandra} (Fig.\ref{fig:Resolve_zero} and \ref{fig:Resolve_phi}).

{\it XRISM}/Resolve has larger a effective area than {\it Chandra}/HEG at high energies band. 
Thus we will access to lines of H- and He-like Ni K$\alpha$, and those of Fe K$\beta$ and even K$\gamma$.
We measure ion columns, Doppler width and blueshift by {\sc kabs} from simulated observation (Tab.\ref{tab:vt_zero} and \ref{tab:vt_vphi}). 
In both simulations, the columns measured from Fe {\sc xxvi} K$\alpha$ and K$\beta$ are significantly different, clearly showing the absorption line saturation in K$\alpha$. 
Thus we can measure ion columns more accurately using those K$\beta$ lines. 
%We can identify these saturation and {\it XRISM}/Resolve c

\begin{table}
    \centering
    \caption{Fits to the absorption lines of  {\it XRISM} simulation from model with $v_\text{turb} = 0$ by {\sc kabs}.  
    Errors are the $90\% $ confidence level ($\Delta\chi^2=2.7$).}
    \scalebox{0.725}{
    \begin{tabular}{ccccc}
    \hline 
   ions & lines & $N_\mathrm{ion}~ [10^{18}~\mathrm{cm^{-2}}]$  & $kT~[\text{keV}] ~(\sqrt{2kT/m_\text{ion}}~[\text{km/s}])$ & $z\times 10^{-3}$ \\ 
   \hline
   %Fe \scriptsize{XXVI} & $3.6^{+46}_{-2.4}$ & $< 49$  & $ -0.91^{+0.37}_{-0.52}$\\
   Fe \sc{xxv} & He$\alpha~y, w$ &$2.2^{+4.0}_{-0.6}$ & $3.0^{+0.8}_{-1.6}~ (100^{+10}_{-30})$  & $ -1.55\pm 0.03$\\
                       & He$\beta~y, w$  &$2 \pm 1$ & $3^{+3}_{-1}~(100^{+50}_{-20})$ & $ -1.58^{+0.06}_{-0.07}$\\
                       %&  He$\gamma~w$ & $4.0^{+2.4}_{-1.1}$ & $3.0^{+1.6}_{-1.0}$  & $ -1.67^{+0.06}_{-0.07}$\\
                       &  He$\gamma~y, w$ & $10^{+380}_{-7}$ & $0.73^{+3.0}_{-0.68}~(50^{+100}_{-23})$  & $ -1.78\pm0.1$\\
   Fe \sc{xxvi} & Ly$\alpha~1,2$ & $1.0^{+0.3}_{-0.2} $ & $5.7^{+1.8}_{-1.6}~(140\pm20)$  & $ -1.71\pm 0.04$\\
                        & Ly$\beta~1, 2$ & $1.8 \pm 0.04 $ & $< 15~(< 230)$  & $ -1.6\pm 0.1$\\
   Ni \sc{xxvii} & He$\alpha~y,w$ &$0.3^{+0.2}_{-0.1}$ & $2^{+2}_{-1}~(80^{+40}_{-20})$  & $ -1.60 \pm 0.06$\\
   Ni \sc{xxviii} & Ly$\alpha~1,2$ &$0.1^{1.3}_{-0.05}$ & $< 22 ~(< 270)$  & $ -1.9\pm 0.2$\\
   %Si \scriptsize{XIV} & $0.052^{+0.040}_{-0.019}$ & $<7.5$  & $ -1.2^{+0.36}_{-0.11}$\\
   \hline 
    \end{tabular}
    }
    \label{tab:vt_zero}
\end{table}

\begin{table}
    \centering
    \caption{Fits to the absorption lines of  {\it XRISM} simulation from model with $v_\text{turb} = v_\phi$ by {\sc kabs}.  Errors are the $90\% $ confidence level ($\Delta\chi^2=2.7$).}
    \scalebox{0.725}{
    \begin{tabular}{ccccc}
    \hline 
   ions & lines & $N_\mathrm{ion}~ [10^{18}~\mathrm{cm^{-2}}]$  & $kT~[\text{keV}]~(\sqrt{2kT/m_\text{ion}}~[\text{km/s}])$ & $z\times 10^{-3}$ \\ 
   \hline
   Fe \sc{xxv} & He$\alpha~y, w$ &$1.7^{+0.3}_{-0.4}$ & $5.1\pm 0.9~(130\pm10)$  & $ -1.77\pm 0.03$\\
                       & He$\beta~y, w$  &$1.9^{+0.4}_{-0.3}$ & $7^{+3}_{-2}~(160^{+30}_{-20})$  & $ -1.73\pm 0.07$\\
                       &  He$\gamma~y, w$ & $2.1\pm 0.5$ & $7^{+7}_{-4}~(160^{+80}_{-40})$  & $ -1.7\pm 0.1$\\
   Fe \sc{xxvi} & Ly$\alpha~1,2$ & $1.4^{+0.2}_{-0.1}$ & $12\pm 2~ (200\pm20) $  & $ -1.88\pm 0.04$\\
                        & Ly$\beta~1, 2$ & $2.3 \pm 0.4 $ & $8^{+8}_{-6}~(170^{+80}_{-60}) $  & $ -1.9\pm 0.1$\\
   Ni \sc{xxvii} & He$\alpha~y,w$ & $0.14^{+0.04}_{-0.03}$ & $6^{+6}_{-3}~(140^{+70}_{-35})$  & $ -1.9\pm 0.1 $\\
   Ni \sc{xxviii} & Ly$\alpha~1,2$ & $0.10^{+0.07}_{-0.05}$ & $4.76^{+26}_{-4.73}~(120^{+340}_{-60})$ & $ -2.0^{+0.2}_{-0.3} $\\
   \hline 
    \end{tabular}
    }
    \label{tab:vt_vphi}
\end{table}

\section{Discussion and Conclusions}

We show the state of the art simulations of the neutron star binary system GX13+1.
This object has the largest disc (in gravitational units) of any known system, and is highly luminous at $\sim 0.5L_\text{Edd}$.
Our radiation hydrodynamic code, which includes radiation pressure on electrons and ions (bound free and lines) as well as X-ray heating, predicts an extremely strong thermal-radiative wind,
and the detailed line profiles computed from fully 3D Monte-Carlo radiation transfer through this wind include a substantial contribution from emission and scattering in the wind, as well as absorption.
The resonance K$\alpha$ lines of Fe {\sc xxv} and {\sc xxvi} are strongly saturated at even moderate inclination, so their EWs depend sensitively on the velocity structure. 
Our radiation hydrodynamic simulation includes the self-consistent velocity shear from acceleration along the line of sight, but we also 
consider the possibility of additional isotropic turbulence at a level comparable to the radial velocity of $\sim 500$~km/s or the azimuthal velocity of $\sim 100$~km/s.
Turbulence at the level of the radial velocity is already ruled out by the {\it Chandra}/HEG third order spectra, which clearly show that the observed absorption line profiles are narrower than this value. 
The line profiles instead appear more consistent with no additional turbulence, with an upper limit from the line profile around the level of the typical azimuthal velocity. 
Future observations with {\it XRISM} will reveal these line profiles in much more detail, while simultaneously constraining the less saturated K$\beta$ and Ni K$\alpha$ lines, to uniquely determine the ion columns. 

However, while the Fe K$\alpha$ {\sc xxvi} and {\sc xxv} line widths are well fit at very low levels of turbulence, the line depth of the Fe {\sc xxvi} K$\alpha$ is clearly underestimated.
It is slightly better fit by including turbulence at the level of the azimuthal velocity, but the line depth is still underestimated, and the line width starts to show some tension.
This indicates that our radiation hydrodynamic simulations have slightly lower ion column in {\sc xxvi} than required by the data.
This lack of column could be due to an underestimate of the ionisation state of gas in the wind
as we also slightly overestimate the column  density of Fe {\sc xxv} (green in Fig.\ref{fig:GX13p1_column}).
Our models do not include the effect of the scattered flux on the ion populations in the radiation hydrodynamic simulation and ionisation calculation.
The effect of the scattered flux is larger in regions where the wind has large column density i.e. at high inclination angles.
Including this additional flux could shift some of the Fe {\sc xxv} to {\sc xxvi} as observed. 
This would shift the Fe {\sc xxv} ion peak to slightly larger radii, 
so give slightly lower velocity in this ion which would also match better to the observed velocity structure.

The column/velocity mismatch could also arise from the subgrid physics of the shadow cast by the X-ray heated atmosphere above the inner disc.
Our models use the analytic estimates of \citet{Begelman1983b} to calculate the radius at which the outer disc rises above this shadow so it can be directly illuminated.
This approach does not include the effect of radiation pressure, 
so is likely not very accurate at these high luminosities.
Nonetheless, the existence of the inner atmosphere will attenuate the illumination of the disc surface to some extent,
and this prevents the large increase in wind column as $L\to L_\text{Edd}$ which was predicted by \cite{Done2018}.
There could also be an additional shadow as $L\to L_\text{Edd}$ from the inner disc itself as it may puff up as it approaches the slim disc regime \citep{Abramowicz1988}.
Understanding the detailed effect of radiation pressure is likely very important in predicting the absorption lines seen in GRS~1915+105 \citep{Neilsen2009} and in the very anomalous wind of GRO J1655-40 \citep{Miller2006}.
 
Despite these tensions, overall the predicted thermal-radiative wind captures most of the observed behaviour of the source.
This is impressive as the model does not include any non-axisymmetic effects 
such as the radiation pressure warping of the outer disc \citep{Ogilvie2001} and/or the effect of the stream impact. 
The observed line widths strongly limit the velocity shear in the wind region to much less than the observed outflow velocity,
and future high resolution observations with {\it XRISM} will reveal this structure in much more detail.
Nonetheless, it is already clear that thermal-radiative winds can explain the features seen, which means that there is no requirement for magnetic winds.
This adds to the growing weight of evidence that the LMXB winds are thermal-radiative rather than magnetically driven.

%%%%%%%%%%%%%%%%%%%%%%%%%%%%%%%%%%%%%%%%%%%%%%%%%%

%%%%%%%%%%%%%%%%% BODY OF PAPER %%%%%%%%%%%%%%%%
%%

%\input{tomaru_hydro}

\section*{Acknowledgements}

This work supported by JSPS KAKENHI Grant Number JP 19J13373 (RT), 
Society for the Promotion of Science Grant-in-Aid for Scientific Research (A) (17H01102 KO; 16H02170 TT), 
Scientific Research (C) (16K05309 KO; 18K03710 KO), 
and Scientific Research on Innovative Areas (18H04592 KO; 18H05463 TT).
This research is also supported by the Ministry of Education, Culture, Sports, Science and Technology of Japan as "Priority Issue on Post-K computer"(Elucidation of the Fundamental Laws and Evolution of the Universe) and JICFuS. 
RT acknowledges the support by JSPS Overseas Challenge Program for Young Resarchers.
CD acknowledges the Science and Technology Facilities Council (STFC) through grant ST/P000541/1, and visitor support from Kavli IPMU  supported in part by the National Science Foundation under Grant No. NSF PHY-1748958.
Numerical computations were in part carried out on Cray XC50 at Center for Computational Astrophysics, National Astronomical Observatory of Japan.

\section*{Data Availability}
X-ray data underlying this article are available at NASA's HEASARC archive (https://heasarc.gsfc.nasa.gov/cgi-bin/W3Browse/w3browse.pl).
Our simulation data underlying this article will be shared on reasonable request to the corresponding author, subject to considerations of intellectual property law.

%%%%%%%%%%%%%%%%%%%%%%%%%%%%%%%%%%%%%%%%%%%%%%%%%%



\bibliographystyle{mnras}
\bibliography{library} % if your bibtex file is called example.bib

%\begin{thebibliography}{99}
%\end{thebibliography}

%%%%%%%%%%%%%%%%%%%%%%%%%%%%%%%%%%%%%%%%%%%%%%%%%%

%%%%%%%%%%%%%%%%% APPENICES %%%%%%%%%%%%%%%%%%%%%

%%%%%%%%%%%%%%%%%%%%%%%%%%%%%%%%%%%%%%%%%%%%%%%%%%

% Don't change these lines
\bsp	% typesetting comment
\label{lastpage}
\end{document}